\documentclass[showpacs  ,secnumarabic, nobibnotes,  nofootinbib, aps,
prd]{revtex4}%
\usepackage{amsmath}
\usepackage{amssymb}
\usepackage{bm}
\usepackage{graphicx}
\usepackage{amsfonts}%
\providecommand{\U}[1]{\protect\rule{.1in}{.1in}}

\begin{document}
\title{Proton spin in leading order of the covariant approach }
\author{Petr Zavada}
\email{zavada@fzu.cz}
\affiliation{Institute of Physics AS CR, Na Slovance 2, CZ-182 21 Prague 8, Czech Republic}

\begin{abstract}
We study the covariant version of the quark-parton model, in which the general
rules of the angular momentum composition are accurately taken into account.
We demonstrate how these rules affect the relativistic interplay between the
quark spins and orbital angular momenta, which collectively contribute to the
proton spin. The spin structure functions $g_{1}$ and $g_{2}$ corresponding to
the many-quark state $J=1/2$ are studied and it is shown they satisfy
constraints and relations well compatible with the available experimental data
including proton spin content $\Delta\Sigma\lesssim1/3$. The suggested Lorentz
invariant 3D approach for calculation of the structure functions is compared
with the approach based on the conventional collinear parton model.

\end{abstract}

\maketitle

\section{Introduction}

The question of correct interpretation and quantitative explanation of the low
value $\Delta\Sigma$ denoting the contribution of spins of quarks to the
proton spin remains still open. Information on the present status of the
understanding of this well known puzzle can be found in the recent review
articles
\cite{Aidala:2012mv,Myhrer:2009uq,Burkardt:2008jw,Barone:2010zz,Kuhn:2008sy}.
It is believed that an important step to the solution of this problem can be a
better understanding of the role of the quark orbital angular momentum (OAM).
In our previous studies
\cite{Zavada:2007ww,Zavada:2002uz,Zavada:2001bq,Efremov:2010cy} we have
suggested the effect of the OAM, if calculated with the help of a covariant
quark-parton model (CQM), can be quite significant. In the present paper we
aim to further develop and extend the study of a common role of the spin and
OAM of quarks. In this connection we reformulate the CQM in terms of the
spinor spherical harmonics - instead of the plane-wave spinors.

In Sec. \ref{model} we summarize the general construction of the CQM and make
a comparison with the conventional parton model. In Sec. \ref{eigenstatesAM}
we discuss in more detail the eigenstates of angular momentum (AM) represented
by the spinor spherical harmonics. Special attention is paid to the
many-particle states resulting from multiple AM composition giving the total
angular momentum $J=1/2$ (i.e. composition of spins and OAMs of all involved
particles). In a next step (Sec. \ref{D&SFs}) these states are used as an
input for calculating of related polarized distribution and structure
functions (SFs) in the general manifestly covariant framework. The same states
are used for definition of the proton state in Sec. \ref{pss}, where it is
shown what sum rules the related SFs satisfy and in particular what can be
predicted for the proton spin content. At the same time the results are
compared with the available experimental data. The last section (Sec.
\ref{summary}) is devoted to the summary of obtained results and concluding
remarks. The Appendices contain some details of the calculations and
supplementary results.

In general the composition of AMs tends to generate rather complicated
expressions for the related matrix elements. That is why we have used the
Wolfram Mathematica (WM) \cite{wolfram} to get or verify some relations and to
simplify obtained expressions. In fact in some cases we used the WM instead of
a rigorous analytic proof, which can be done later in a separate study. Such
expressions, where the use of WM was essential, are provided by the note
\textit{obtained with WM}.

\section{Model}

\label{model}The basis of our present approach is the CQM, which has been
studied earlier
\cite{Zavada:2011cv,Zavada:2009sk,Zavada:2007ww,Zavada:2002uz,Zavada:2001bq,Zavada:1996kp,Efremov:2010cy,Efremov:2010mt,Efremov:2009ze,Efremov:2004tz}%
. This model was motivated by the parton model suggested by R. Feynman
\cite{fey}. The important differences between them will be explained below,
but the main postulates, which are common for the CQM and the conventional
parton model can be formulated as follows:

\textit{i)} The deep inelastic scattering (DIS) can be (in a leading order)
described as an incoherent superposition of interactions of a probing lepton
with the individual effectively free quarks (partons) inside the nucleon. The
lepton-quark scattering is described by the one-photon exchange diagram, from
which the corresponding quark tensor is obtained. It means that the
photon-quark interaction is assumed to be quasi-instantaneous and that the
final state interactions are ignored.

\textit{ii)} The kinematical degrees of freedom of the quarks inside the
nucleon are described by a set of probabilistic distribution functions.
Integration of the quark tensors with the corresponding distributions gives
the hadronic tensor, from which the related SFs are obtained.

In the conventional model this picture is assumed only in the frame, where the
proton is fast moving. The paradigm of the CQM is different, we assume that
during the interaction at sufficiently high $Q^{2}$ the quark can be in a
leading order neglecting the QCD corrections considered effectively free in
\textit{any} reference frame. The argument is as follows. The space-time
dimensions $\Delta\lambda\times\Delta\tau$ of the quark vicinity where the
interaction takes place is defined by the photon momentum squared $q^{2}%
=q_{0}^{2}-\mathbf{q}^{2}=-Q^{2}$ and Bjorken variable $x=Q^{2}/\left(
2P\cdot q\right)  $. In the proton rest frame using the standard notation
$q_{0}=\nu$ we have%
\begin{equation}
\mathbf{q}_{R}^{2}=Q^{2}+\nu^{2}=Q^{2}\left(  1+\frac{Q^{2}}{\left(
2Mx\right)  ^{2}}\right)  , \label{rm2}%
\end{equation}
which implies%
\begin{equation}
\left\vert \mathbf{q}_{R}\right\vert \gtrsim\nu=\frac{Q^{2}}{2Mx}\geq
\frac{Q^{2}}{2M}, \label{rm3}%
\end{equation}
so the space-time domain in the rest frame, where the interaction takes place,
is limited:
\begin{equation}
\Delta\lambda\lesssim\Delta\tau\approx\frac{2Mx}{Q^{2}}. \label{rm4}%
\end{equation}
Let us mention, a space-time picture of photon and neutrino scattering was
discussed already in \cite{ioffe}. The last relation means that the quark at
sufficiently large $Q^{2}$, due to the effect of asymptotic freedom, must
behave during interaction with probing lepton as if it was free. For example,
for $x=0.3$ and $Q^{2}=10$ GeV$^{2}$ we have $\Delta\lambda\lesssim\Delta
\tau\approx0.06$ fm ($1$ GeV$^{-1}=0.197$ fm). The limited extent of the
domain prevents the quark from any interaction with the rest of nucleon,
absence of interaction is synonym for freedom. Apparently, this argument is
valid in any reference frame as it is illustrated in Fig. \ref{fgr1}, where
the light cone domain $\Delta\tau=0.25$ fm in the nucleon of radius $R_{n}=$
$0.8$ fm is displayed for different Lorentz boosts: \begin{figure}[ptb]
\includegraphics[width=12cm]{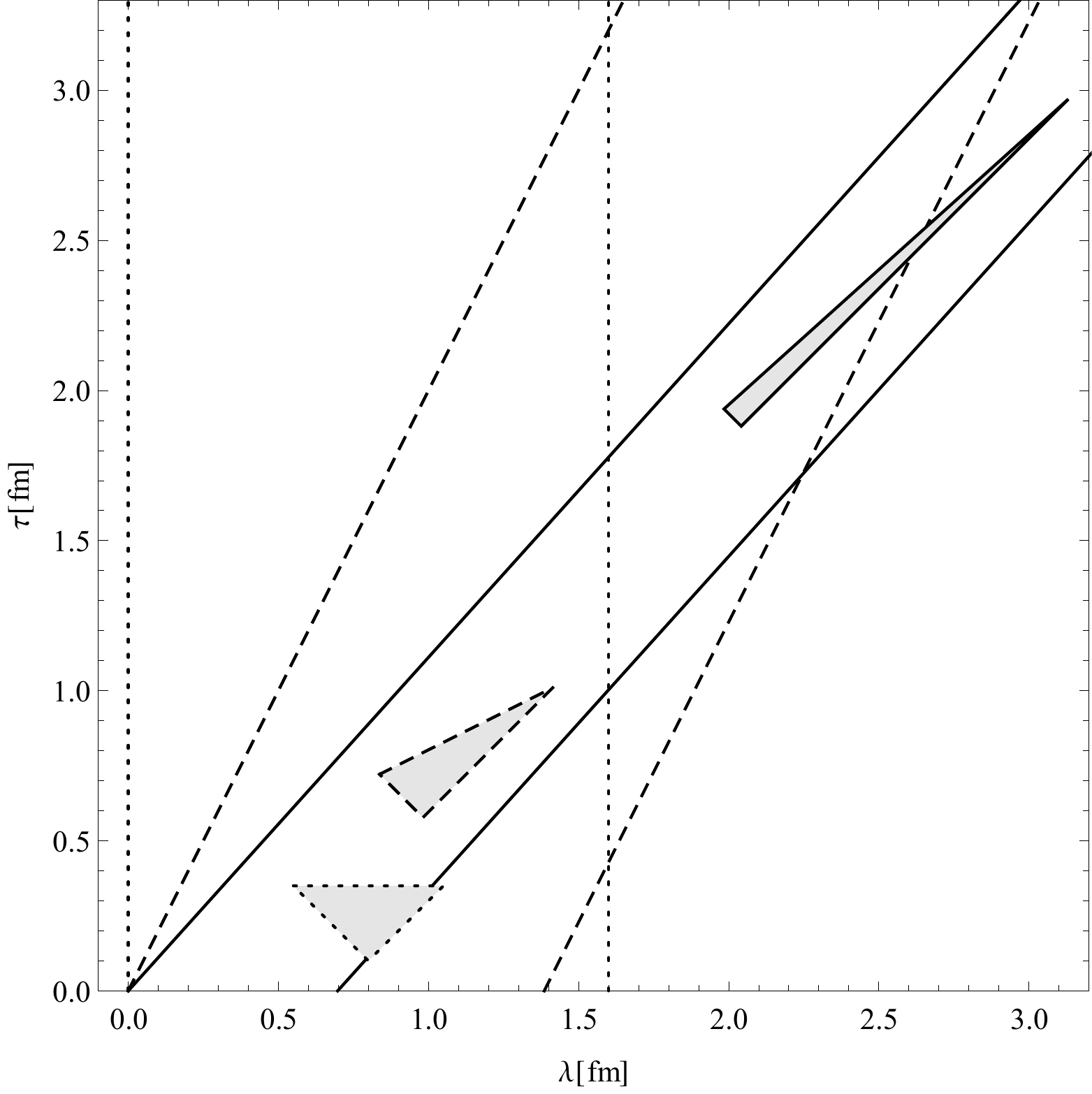}\caption{ The space-time domain of the
photon momentum transfer to the quark in different Lorentz frames. The
different styles of lines and triangles represent the proton boundary and the
domain for: rest frame, $\beta=0$ \textit{(dotted)}, $\beta=0.5$
\textit{(dashed)}, $\beta=0.9$ \textit{(solid)}. Note that Lorentz boosts does
not change the area of the domain $\Delta\lambda\times\Delta\tau$.}%
\label{fgr1}%
\end{figure}%
\begin{equation}
\lambda(\beta)=\frac{\lambda_{0}+\beta\tau_{0}}{\sqrt{1-\beta^{2}}},\qquad
\tau(\beta)=\frac{\tau_{0}+\beta\lambda_{0}}{\sqrt{1-\beta^{2}}}. \label{rm9}%
\end{equation}
The figure also illustrates that in the frame where the nucleon is fast
moving, the time is dilated and the lengths are Lorentz-contracted (nucleon
and the light cone domain are made flatter). It means intrinsic motion is
slowed down and the interaction takes correspondingly longer time. In fact we
work with the approximation%
\begin{equation}
\Delta\tau\ll\Delta\tau_{QCD}, \label{rm10}%
\end{equation}
where $\Delta\tau_{QCD}$ is characteristic time of the QCD process
accompanying the photon absorption. The Lorentz time dilatation%
\begin{equation}
\Delta T(\beta)=\frac{\Delta T_{0}}{\sqrt{1-\beta^{2}}} \label{rm11}%
\end{equation}
is universal, so we assume the relation (\ref{rm10}) to be valid in any
boosted frame. In other words we assume the characteristic time $\Delta
\tau_{QCD}$ have a good sense in any reference frame even if we are not able
to transform the QCD correction itself, this correction is calculable by the
perturbative QCD only in the infinite momentum frame (IMF). This is essence of
our covariant leading order approach. Of course, the large but finite $Q^{2}%
$\ gives a room for interaction with a limited neighborhood of gluons and sea
quarks. Then the role of effectively free quarks is played by these "dressed"
quarks inside the corresponding domain. The shape of the domain changes with
Lorentz boost, but the physics inside remains the same. Let us point out the
CQM does not aim to describe the complete nucleon dynamic structure, but only
a very short time interval $\Delta\tau$ during DIS. The aim is to describe and
interpret the DIS data. For a fixed $Q^{2}$ the CQM approach represents a
picture of the nucleon with a set of quarks taken in a thin time slice
(limited space-time domain). Quantitatively, the $Q^{2}-$dependence of this
image is controlled by the QCD. We assume that the approximation of quarks by
the free waves in this limited space-time domain is acceptable for description
of DIS regardless of the reference frame.

Let us remark, the argument often used in favor of conventional approach based
on the IMF is as follows \cite{imf}: ... \textit{Additionally, the hadron is
in a reference frame where it has infinite momentum --- a valid approximation
at high energies. Thus, parton motion is slowed by time dilation, and the
hadron charge distribution is Lorentz-contracted, so incoming particles will
be scattered "instantaneously and incoherently"...} In our opinion one should
add, not only parton motion but also energy transfer are slowed down. One can
hardly speak about "instantaneous" scattering without reference to the
invariant scale $Q^{2},$ cf. Fig. \ref{fgr1}. In our opinion, if $Q^{2}$ is
sufficiently large to ensure the scattering is "instantaneous" in the IMF,
then the same statement holds for any boosted frame.

In the framework of conventional approach there exist algorithms for the
$Q^{2}$ evolution, i.e. knowledge of a distribution function (PDF) at some
initial scale $Q_{0}^{2}$ allows us to predict it at another scale. Such
algorithm is not presently known for covariant approach. But in CQM the
knowledge of one PDF at some scale allows us to predict some another PDF at
the same scale. The set of corresponding rules involves also transverse
momentum distribution functions (TMD)
\cite{Efremov:2010cy,Efremov:2010mt,Efremov:2009ze,Efremov:2004tz}. Therefore,
from phenomenological point of view, there is a complementarity between both approaches.

The main practical difference between the approaches is in the input
probabilistic distribution functions. The conventional IMF distributions, due
to simplified one dimensional kinematics, are easier for handling, e.g. their
relation to the SFs is extremely simple. On the other hand the CQM
distributions, reflecting 3D kinematics of quarks and depending on $Q^{2}$,
require a more complicated but feasible construction to obtain SFs. However
the CQM in the limit of static quarks is equivalent to the collinear approach,
see Appendix A in \cite{Zavada:2007ww}. The difference in predictions
following from both the approaches is significant particularly for the
polarized SFs and that is why in this paper we pay attention mainly to the
polarized DIS.

\section{Eigenstates of angular momentum}

\label{eigenstatesAM}The solutions of free Dirac equation represented by
eigenstates of the total angular momentum (AM) with quantum numbers $j,j_{z}$
are the spinor spherical harmonics \cite{bdtm,lali, bie}, which in the
\textit{momentum representation} reads:%
\begin{equation}
\left\vert j,j_{z}\right\rangle =\Phi_{jl_{p}j_{z}}\left(  \mathbf{\omega
}\right)  =\frac{1}{\sqrt{2\epsilon}}\left(
\begin{array}
[c]{c}%
\sqrt{\epsilon+m}\Omega_{jl_{p}j_{z}}\left(  \mathbf{\omega}\right)  \\
-\sqrt{\epsilon-m}\Omega_{j\lambda_{p}j_{z}}\left(  \mathbf{\omega}\right)
\end{array}
\right)  ,\label{rs1}%
\end{equation}
where $\mathbf{\omega}$ represents the polar and azimuthal angles
($\theta,\varphi$) of the momentum $\mathbf{p}$ with respect\ to the
quantization axis $z,$ $l_{p}=j\pm1/2,\ \lambda_{p}=2j-l_{p}$ ($l_{p}$ defines
the parity), energy $\epsilon=\sqrt{\mathbf{p}^{2}+m^{2}}$ and
\begin{align}
\Omega_{jl_{p}j_{z}}\left(  \mathbf{\omega}\right)   &  =\left(
\begin{array}
[c]{c}%
\sqrt{\frac{j+j_{z}}{2j}}Y_{l_{p},j_{z}-1/2}\left(  \mathbf{\omega}\right)  \\
\sqrt{\frac{j-j_{z}}{2j}}Y_{l_{p},j_{z}+1/2}\left(  \mathbf{\omega}\right)
\end{array}
\right)  ;\quad l_{p}=j-\frac{1}{2},\label{rs1e}\\
\Omega_{jl_{p}j_{z}}\left(  \mathbf{\omega}\right)   &  =\left(
\begin{array}
[c]{c}%
-\sqrt{\frac{j-j_{z}+1}{2j+2}}Y_{l_{p},j_{z}-1/2}\left(  \mathbf{\omega
}\right)  \\
\sqrt{\frac{j+j_{z}+1}{2j+2}}Y_{l_{p},j_{z}+1/2}\left(  \mathbf{\omega
}\right)
\end{array}
\right)  ;\quad l_{p}=j+\frac{1}{2}.\nonumber
\end{align}
The structure of these spinors follows from composition of the orbital and
spin components with the use of corresponding Clebsch-Gordan coefficients:%
\begin{equation}
\Omega_{jl_{p}j_{z}}\left(  \mathbf{\omega}\right)  =\left\langle l_{p}%
,j_{z}-1/2,1/2,1/2\left\vert j,j_{z}\right.  \right\rangle Y_{l_{p},j_{z}%
-1/2}\left(  \mathbf{\omega}\right)  \binom{1}{0}+\left\langle l_{p}%
,j_{z}+1/2,1/2,-1/2\left\vert j,j_{z}\right.  \right\rangle Y_{l_{p}%
,j_{z}+1/2}\left(  \mathbf{\omega}\right)  \binom{0}{1}.\label{rs1g}%
\end{equation}
Let us remind that in relativistic case the quantum numbers of spin and OAM
are not conserved separately, but only the total AM $j$ and its projection
$j_{z}=s_{z}+l_{z}$ can be conserved. The complete wave function reads:%
\begin{equation}
\Psi_{jl_{p}j_{z}}\left(  \epsilon,\mathbf{\omega}\right)  =\phi_{j}\left(
\epsilon\right)  \Phi_{jl_{p}j_{z}}\left(  \mathbf{\omega}\right)
.\label{rs1a}%
\end{equation}
The function $\phi_{j}\left(  \epsilon\right)  $\ or its equivalent
representation (\ref{rs1f}) is amplitude of probability that the fermion has
energy \ $\epsilon$. In fact the main results obtained in this paper depend
only on the probability distribution $a_{j}^{\ast}\left(  \epsilon\right)
a_{j}\left(  \epsilon\right)  $ via the parameters (\ref{rs8b}) and the
functions $\mathrm{u}\left(  \epsilon\right)  ,\mathrm{v}\left(
\epsilon\right)  $ defining the general spin vector (\ref{rs47c}). The spinors
(\ref{rs1}) are normalized as%
\begin{equation}
\int\Phi_{j^{\prime}l_{p}^{\prime}j_{z}^{\prime}}^{+}\left(  \mathbf{\omega
}\right)  \Phi_{jl_{p}j_{z}}\left(  \mathbf{\omega}\right)  d\mathbf{\omega
}=\delta_{j^{\prime}j}\delta_{l_{p}^{\prime}l_{p}}\delta_{j_{z}^{\prime}j_{z}%
},\label{rs1b}%
\end{equation}
where $d\mathbf{\omega=}d\cos\theta\ d\varphi.$ Then the normalization%
\begin{equation}
\int\Psi_{j^{\prime}l_{p}^{\prime}j_{z}^{\prime}}^{+}\left(  \epsilon
,\mathbf{\omega}\right)  \Psi_{jl_{p}j_{z}}\left(  \epsilon,\mathbf{\omega
}\right)  d^{3}\mathbf{p}=\delta_{j^{\prime}j}\delta_{l_{p}^{\prime}l_{p}%
}\delta_{j_{z}^{\prime}j_{z}}\label{rs1c}%
\end{equation}
implies the condition for the amplitude $\phi_{j}:$
\begin{equation}
\int\phi_{j}^{\ast}\left(  \epsilon\right)  \phi_{j}\left(  \epsilon\right)
p^{2}dp=1.\label{rs1d}%
\end{equation}
In the next discussion it will be convenient to use also the alternative
representation, which differs in normalization:
\begin{equation}
a_{j}\left(  \epsilon\right)  =\frac{\phi_{j}\left(  \epsilon\right)  }%
{2\sqrt{\pi}};\qquad\int a_{j}^{\ast}\left(  \epsilon\right)  a_{j}\left(
\epsilon\right)  d^{3}\mathbf{p}=1.\label{rs1f}%
\end{equation}

\subsection{Angular moments of one-fermion states}

A few examples of the probability distribution corresponding to the states
(\ref{rs1})%
\begin{equation}
P_{j,j_{z}}\left(  \mathbf{\omega}\right)  =\Phi_{jl_{p}j_{z}}^{+}\left(
\mathbf{\omega}\right)  \Phi_{jl_{p}j_{z}}\left(  \mathbf{\omega}\right)
;\qquad\int P_{j,j_{z}}\left(  \mathbf{\omega}\right)  d\mathbf{\omega}=1,
\label{rs2}%
\end{equation}
are given in the first panel of Tab. \ref{tb1}. Let us remark this
distribution does not depend on the parameters $\varphi$\ and $l_{p}=j\pm1/2$.
\begin{table}[ptb]
\begin{center}%
\begin{tabular}
[c]{c|c|c|c|}%
$j,j_{z}$ & $P_{j,j_{z}}(\omega)$ & $H_{j,j-1/2,j_{z}}(\omega)$ &
$H_{j,j+1/2,j_{z}}(\omega)$\\\hline
$\frac{1}{2},\frac{1}{2}$ & $1$ & $1$ & $\cos2\theta$\\
$\frac{3}{2},\frac{3}{2}$ & $\frac{3-3\cos2\theta}{4}$ & $\frac{3-3\cos
2\theta}{4}$ & $\frac{-3+6\cos2\theta-3\cos4\theta}{8}$\\
$\frac{3}{2},\frac{1}{2}$ & $\frac{5+3\cos2\theta}{4}$ & $\frac{3+5\cos
2\theta}{4}$ & $\frac{1+6\cos2\theta+9\cos4\theta}{8}$\\
$\frac{5}{2},\frac{5}{2}$ & $\frac{45-60\cos2\theta+15\cos4\theta}{64}$ &
$\frac{45-60\cos2\theta+15\cos4\theta}{64}$ & $\frac{-60+105\cos2\theta
-60\cos4\theta+15\cos6\theta}{128}$\\
$\frac{5}{2},\frac{3}{2}$ & $\frac{57-12\cos2\theta-45\cos4\theta}{64}$ &
$\frac{39+12\cos2\theta-51\cos4\theta}{64}$ & $\frac{-36+51\cos2\theta
+60\cos4\theta-75\cos6\theta}{128}$\\
$\frac{5}{2},\frac{1}{2}$ & $\frac{45+36\cos2\theta+15\cos4\theta}{32}$ &
$\frac{21+36\cos2\theta+39\cos4\theta}{32}$ & $\frac{12+45\cos2\theta
+60\cos4\theta+75\cos6\theta}{64}$%
\end{tabular}
\end{center}
\caption{The examples of the distributions (\ref{rs2}) and (\ref{rs38}). The
common factor $1/4\pi$ is omitted. For opposite $j_{z}$ it holds:
$P_{j,-j_{z}}=P_{j,j_{z}}$ and $H_{j,l,-j_{z}}=-H_{j,l,j_{z}}$.}%
\label{tb1}%
\end{table}The lowest value $j=1/2$ generates rotational symmetry of the
probability distribution, for higher $j=3/2,5/2,...$ the distribution has
axial symmetry only. The states (\ref{rs1}) are not eigenstates of spin and
OAM, nevertheless one can always calculate the mean values of corresponding
operators%
\begin{equation}
s_{z}=\frac{1}{2}\left(
\begin{array}
[c]{cc}%
\sigma_{z} & 0\\
0 & \sigma_{z}%
\end{array}
\right)  ,\qquad l_{z}=-i\left(  p_{x}\frac{\partial}{\partial p_{y}}%
-p_{y}\frac{\partial}{\partial p_{x}}\right)  . \label{rs3}%
\end{equation}
The related matrix elements are given by the relations (\textit{obtained with
WM}):%
\begin{align}
\left\langle s_{z}\right\rangle _{j,j_{z}}  &  =\int\Phi_{jl_{p}j_{z}}%
^{+}s_{z}\Phi_{jl_{p}j_{z}}d\mathbf{\omega}=\frac{1+\left(  2j+1\right)  \mu
}{4j\left(  j+1\right)  }j_{z},\label{rs4}\\
\left\langle l_{z}\right\rangle _{j,j_{z}}  &  =\int\Phi_{jl_{p}j_{z}}%
^{+}l_{z}\Phi_{jl_{p}j_{z}}d\mathbf{\omega}=\left(  1-\frac{1+\left(
2j+1\right)  \mu}{4j\left(  j+1\right)  }\right)  j_{z},\nonumber
\end{align}
in which we have denoted
\begin{equation}
\mu=\pm\frac{m}{\epsilon}, \label{rs4a}%
\end{equation}
where the sign $\left(  \pm\right)  $ corresponds to $l_{p}=j\mp1/2$. The
relations imply that in the non-relativistic limit, when $\mu\simeq\pm1,$ we
get for both signs $\left(  \pm\right)  $ \ correspondingly%
\begin{equation}
\left\langle s_{z}\right\rangle _{j,j_{z}}=%
\genfrac{\{}{\}}{0pt}{}{\frac{j_{z}}{2j}}{\frac{-j_{z}}{2\left(  j+1\right)
}}%
,\qquad\left\langle l_{z}\right\rangle _{j,j_{z}}=%
\genfrac{\{}{\}}{0pt}{}{\left(  1-\frac{1}{2j}\right)  j_{z}}{\left(
1+\frac{1}{2\left(  j+1\right)  }\right)  j_{z}}
\label{rs6}%
\end{equation}
and in the relativistic case, when $\mu\rightarrow0,$ we have%
\begin{equation}
\left\langle s_{z}\right\rangle _{j,j_{z}}=\frac{j_{z}}{4j\left(  j+1\right)
},\qquad\left\langle l_{z}\right\rangle _{j,j_{z}}=\left(  1-\frac
{1}{4j\left(  j+1\right)  }\right)  j_{z}. \label{rs7}%
\end{equation}
The last two relations imply%
\begin{equation}
\left\vert \left\langle s_{z}\right\rangle _{j,j_{z}}\right\vert \leq\frac
{1}{4\left(  j+1\right)  }\leq\frac{1}{6},\qquad\frac{\left\vert \left\langle
s_{z}\right\rangle _{j,j_{z}}\right\vert }{\left\vert \left\langle
l_{z}\right\rangle _{j,j_{z}}\right\vert }\leq\frac{1}{4j^{2}+4j-1}\leq
\frac{1}{2}. \label{rs8}%
\end{equation}
Apparently the bounds obtained in \cite{Zavada:2007ww} \ represent a special
case of these inequalities. For the complete wave function (\ref{rs1a}) the
relations (\ref{rs4}) are modified as
\begin{align}
\left\langle \left\langle s_{z}\right\rangle \right\rangle _{j,j_{z}}  &
=\int\Psi_{_{jl_{p}j_{z}}}^{+}s_{z}\Psi_{jl_{p}j_{z}}d^{3}\mathbf{p}%
=\frac{1+\left(  2j+1\right)  \left\langle \mu_{j}\right\rangle }{4j\left(
j+1\right)  }j_{z},\label{rs8a}\\
\left\langle \left\langle l_{z}\right\rangle \right\rangle _{j,j_{z}}  &
=\int\Psi_{_{jl_{p}j_{z}}}^{+}l_{z}\Psi_{jl_{p}j_{z}}d^{3}\mathbf{p}=\left(
1-\frac{1+\left(  2j+1\right)  \left\langle \mu_{j}\right\rangle }{4j\left(
j+1\right)  }\right)  j_{z},\nonumber
\end{align}
where%
\begin{equation}
\left\langle \mu_{j}\right\rangle =\pm\int a_{j}^{\ast}\left(  \epsilon
\right)  a_{j}\left(  \epsilon\right)  \frac{m}{\epsilon}d^{3}\mathbf{p}%
,\qquad\left\vert \left\langle \mu_{j}\right\rangle \right\vert \leq1.
\label{rs8b}%
\end{equation}

\subsection{Many-fermion states}

The system of fermions (or arbitrary particles) generating the state with
quantum numbers $J,J_{z}$ can be represented by the combination of
one-particle states. For example the pair of states $j_{1},j_{2}$ can generate
the states%
\begin{align}
\left\vert (j_{1},j_{2})J,J_{z}\right\rangle  &  =\sum_{j_{z1}=-j_{1}}^{j_{1}%
}\sum_{j_{z2}=-j_{2}}^{j_{2}}\left\langle j_{1},j_{z1},j_{2},j_{z2}\left\vert
J,J_{z}\right.  \right\rangle \left\vert j_{1},j_{z1}\right\rangle \left\vert
j_{2},j_{z2}\right\rangle ;\label{rs9}\\
j_{z1}+j_{z2}  &  =J_{z},\qquad\qquad\qquad\left\vert j_{1}-j_{2}\right\vert
\leq J\leq j_{1}+j_{2}, \label{rs9a}%
\end{align}
where $\left\langle j_{1},j_{z1},j_{2},j_{z2}\left\vert J,J_{z}\right.
\right\rangle $ are Clebsch-Gordan coefficients, which are non-zero iff the
conditions (\ref{rs9a}) are satisfied. In this way one can repeat the
composition and obtain the many-particle eigenstates of resulting $J,J_{z}$%
\begin{equation}
\left\vert (j_{1},j_{2},...j_{n})_{c}J,J_{z}\right\rangle =\sum_{j_{z1}%
=-j_{1}}^{j_{1}}\sum_{j_{z2}=-j_{2}}^{j_{2}}...\sum_{j_{zn}=-j_{n}}^{j_{n}%
}c_{j}\left\vert j_{1},j_{z1}\right\rangle \left\vert j_{2},j_{z2}%
\right\rangle ...\left\vert j_{n},j_{zn}\right\rangle , \label{rs10}%
\end{equation}
where the coefficients $c_{j}$ consist of the Clebsch-Gordan coefficients%
\begin{equation}
c_{j}=\left\langle j_{1},j_{z1},j_{2},j_{z2}\left\vert J_{3},J_{3z}\right.
\right\rangle \left\langle J_{3},J_{z3},j_{3},j_{z3}\left\vert J_{4}%
,J_{z4}\right.  \right\rangle ...\left\langle J_{n},J_{zn},j_{n}%
,j_{zn}\left\vert J,J_{z}\right.  \right\rangle . \label{rs10b}%
\end{equation}
Let us remark the set $j_{1},j_{2},..j_{n}$ does not define the resulting
state unambiguously. The result depends on the pattern of \ their composition,
e.g. \newline%
\begin{equation}
(((j_{1}\oplus j_{2})_{J_{1}}\oplus j_{3})_{J_{2}}\oplus j_{4})_{J}%
,\qquad(((j_{1}\oplus j_{2})_{J_{1}}\oplus(j_{3}\oplus j_{4})_{J_{2}})_{J_{3}%
}\oplus j_{5})_{J}, \label{10d}%
\end{equation}
where $J_{k}$ represent intermediate AMs corresponding to the steps of
composition:%
\begin{equation}
j_{1}\oplus j_{2}=J_{1},\qquad J_{1}\oplus j_{3}=J_{2},\qquad J_{2}\oplus
j_{4}=J. \label{10e}%
\end{equation}
Each binary composition "$\oplus$"\ is defined by Eq. (\ref{rs9}). Different
composition patterns are in (\ref{rs10}) symbolically expressed by the
subscript $c$. Apparently, the number of patterns increases with $n$ very
rapidly, however in a real scenario with an interaction one can expect their
probabilities will differ. The case $n=3$ will be illustrated in more detail below.

From now we discuss only the composed states with resulting $J=J_{z}=1/2$
($J_{z}=-1/2$ gives the equivalent results). The corresponding $n-$fermion
state ($n$ is odd)%
\begin{equation}
\Phi_{c,1/2,1/2}(\mathbf{\omega}_{1},\mathbf{\omega}_{2},..\mathbf{\omega}%
_{n})=\left\vert (j_{1},j_{2},...j_{n})_{c}1/2,1/2\right\rangle ,
\label{rs10a}%
\end{equation}
or alternatively%
\begin{equation}
\Psi_{c,1/2,1/2}=\phi_{j_{1}}\left(  \epsilon_{1}\right)  \phi_{j_{2}}\left(
\epsilon_{2}\right)  ..\phi_{j_{n}}\left(  \epsilon_{n}\right)  \Phi
_{c,1/2,1/2}(\mathbf{\omega}_{1},\mathbf{\omega}_{2},..\mathbf{\omega}_{n})
\label{rs10c}%
\end{equation}
generate the $n-$dimensional angular distribution
\begin{equation}
P_{c}(\mathbf{\omega}_{1},\mathbf{\omega}_{2},..\mathbf{\omega}_{n}%
)=\Phi_{c,1/2,1/2}^{+}\Phi_{c,1/2,1/2}, \label{rs11}%
\end{equation}
from which the corresponding average one-fermion distributions are obtained as%
\begin{equation}
p_{c,k}(\mathbf{\omega}_{k})=\int P_{c}(\mathbf{\omega}_{1},\mathbf{\omega
}_{2},..\mathbf{\omega}_{n})\prod_{i\neq k}^{n}d\mathbf{\omega}_{i},
\label{rs11a}%
\end{equation}
which gives (\textit{obtained with WM}):%

\begin{equation}
p_{c,k}(\mathbf{\omega})=\frac{1}{4\pi}. \label{RS11G}%
\end{equation}
We have proven this result in Appendix \ref{app1} for one particular
composition pattern. It follows that the distribution%

\begin{equation}
P_{c}(\mathbf{\omega})=\sum_{k=1}^{n}p_{c,k}(\mathbf{\omega})=\frac{n}{4\pi},
\label{rs16b}%
\end{equation}
which is generated by the state (\ref{rs10a}) has rotational symmetry equally
as the distribution $P_{1/2,1/2}$ generated by the one-fermion state in Tab.
\ref{tb1}. Therefore the angular probability distribution $P_{c}%
(\mathbf{\omega})$ related to the state $J=1/2$ has rotational symmetry
regardless of the number of involved particles. Let us remark, this general
rule can be applied also e.g. to the nuclei. It suggests that in a nucleus
$J=1/2$ probability distribution of nucleons, separately of protons and
neutrons, has in the momentum space rotational symmetry. Similar argument is
valid for nucleons in the CQM approach, where the state of partons is
represented by Eqs. (\ref{rs20}) or (\ref{rs22}) below. Spherical symmetry of
probability distribution in momentum space apparently implies spherical
symmetry in coordinate representation.

What can be said about the mean values of the spin and OAM contributions
\begin{gather}
\left\langle \mathbb{S}_{z}\right\rangle _{c,1/2,1/2}=\left\langle
s_{z1}+s_{z2}+...+s_{zn}\right\rangle _{c},\qquad\left\langle \mathbb{L}%
_{z}\right\rangle _{c,1/2,1/2}=\left\langle l_{z1}+l_{z2}+...+l_{zn}%
\right\rangle _{c},\label{rs16}\\
\left\langle \mathbb{S}_{z}\right\rangle _{c,1/2,1/2}+\left\langle
\mathbb{L}_{z}\right\rangle _{c,1/2,1/2}=\frac{1}{2},\nonumber
\end{gather}
corresponding to the state (\ref{rs10a})? In the next we discuss this question
in more detail for the case $n=3,$ which is sufficiently illustrative and will
be of practical importance for our approach.

\subsubsection{Three-fermion states}

\label{3fermions}There are three patterns for composition of\ the three AMs
$j_{a},j_{b},j_{c}$:%
\begin{equation}
(\left(  j_{a}\oplus j_{b}\right)  _{J_{c}}\oplus j_{c})_{1/2};\qquad
abc=123,312,231. \label{ri1}%
\end{equation}
Corresponding states are
\begin{gather}
\Phi_{c,1/2,1/2}(\mathbf{\omega}_{1},\mathbf{\omega}_{2},\mathbf{\omega}%
_{3})=\sum_{j_{z1}=-j_{1}}^{j_{1}}\sum_{j_{z2}=-j_{2}}^{j_{2}}\sum
_{j_{z3}=-j_{3}}^{j_{3}}\left\langle j_{a},j_{za},j_{b},j_{zb}\left\vert
J_{c},J_{zc}\right.  \right\rangle \left\langle J_{c},J_{zc},j_{c}%
,j_{zc}\left\vert 1/2,1/2\right.  \right\rangle \label{rs12}\\
\times\left\vert j_{1},j_{z1}\right\rangle \left\vert j_{2},j_{z2}%
\right\rangle \left\vert j_{3},j_{z3}\right\rangle .\nonumber
\end{gather}
The conditions (\ref{rs9a}) give at most two possibilities for the
intermediate values $J_{c},$ which must satisfy
\begin{equation}
J_{c}=j_{c}\pm1/2,\qquad\left\vert j_{a}-j_{b}\right\vert \leq J_{c}\leq
j_{a}+j_{b}. \label{rs13}%
\end{equation}
At the same time it holds%
\begin{equation}
j_{z1}+j_{z2}+j_{z3}=1/2,\qquad j_{za}+j_{zb}=J_{zc}. \label{rs15}%
\end{equation}
In this way two possible values $J_{c}$ in three patterns (\ref{ri1}) give six
possibilities to create the state (\ref{rs12}). Further, taking into account
two possible values $l_{p}=j\pm1/2$ for each one-fermion state in (\ref{rs12})
and defined by (\ref{rs1}), then in general the total number of generated
three-fermion states is $6\times2^{3}=48$. Due to orthogonality of the terms
in sum (\ref{rs12}) the three-fermion mean values (\ref{rs16}) are calculated
as:
\begin{equation}
\left\langle \mathbb{S}_{z}\right\rangle _{c,1/2,1/2}=\sum_{j_{z1}=-j_{1}%
}^{j_{1}}\sum_{j_{z2}=-j_{2}}^{j_{2}}\sum_{j_{z3}=-j_{3}}^{j_{3}}\left\vert
\left\langle j_{a},j_{za},j_{b},j_{zb}\left\vert J_{c},J_{zc}\right.
\right\rangle \left\langle J_{c},J_{zc},j_{c},j_{zc}\left\vert 1/2,1/2\right.
\right\rangle \right\vert ^{2}\left(  \left\langle s_{za}\right\rangle
+\left\langle s_{zb}\right\rangle +\left\langle s_{zc}\right\rangle \right)
\label{rs17}%
\end{equation}
and similarly for $\left\langle \mathbb{L}_{z}\right\rangle _{c,1/2,1/2}$.
Corresponding one-fermion values $\left\langle s_{z..}\right\rangle $ and
$\left\langle l_{z..}\right\rangle $ are given by the relations (\ref{rs4}).
The results for a set of input values $j_{1},j_{2},j_{3}$ and $l_{pk}%
=j_{k}-1/2$ are listed in Tab. \ref{tb2} and the results corresponding to
remaining sets $l_{pk}=j_{k}\pm1/2$ are similar and differ only in terms
proportional to $\tilde{\mu}$ (\textit{obtained with WM}). Since%
\begin{equation}
\left\langle \mathbb{S}_{z}\right\rangle _{c,1/2,1/2}=-\left\langle
\mathbb{S}_{z}\right\rangle _{c,1/2,-1/2},\qquad\left\langle \mathbb{S}%
_{z}\right\rangle _{c,1/2,\pm1/2}+\left\langle \mathbb{L}_{z}\right\rangle
_{c,1/2,\pm1/2}=\pm1/2, \label{rs18}%
\end{equation}
we present only $\left\langle \mathbb{S}_{z}\right\rangle _{c}\equiv
\left\langle \mathbb{S}_{z}\right\rangle _{c,1/2,1/2}$. \begin{table}[ptb]
\begin{center}%
\begin{tabular}
[c]{ccc|ccc|ccc}%
$j_{1}$ & $j_{2}$ & $j_{3}$ & $\left\langle S_{z}\right\rangle _{3}$ &
$\left\langle S_{z}\right\rangle _{2}$ & $\left\langle S_{z}\right\rangle
_{1}$ & $\left\langle S_{z}\right\rangle _{3}$ & $\left\langle S_{z}%
\right\rangle _{2}$ & $\left\langle S_{z}\right\rangle _{1}$\\\hline
$\frac{1}{2}$ & $\frac{1}{2}$ & $\frac{1}{2}$ & $\frac{1+2\tilde{\mu}}{6}$ &
$\frac{1+2\tilde{\mu}}{6}$ & $\frac{1+2\tilde{\mu}}{6}$ & $\frac{1+2\tilde
{\mu}}{6}$ & $\frac{1+2\tilde{\mu}}{6}$ & $\frac{1+2\tilde{\mu}}{6}$\\
$\frac{3}{2}$ & $\frac{1}{2}$ & $\frac{1}{2}$ & $\times$ & $\times$ &
$\frac{-1}{18}$ & $\frac{-1}{18}$ & $\frac{-1}{18}$ & $\times$\\
$\frac{3}{2}$ & $\frac{3}{2}$ & $\frac{1}{2}$ & $\frac{1+2\tilde{\mu}}{6}$ &
$\frac{1+3\tilde{\mu}}{18}$ & $\frac{1+3\tilde{\mu}}{18}$ & $\frac
{-1+6\tilde{\mu}}{90}$ & $\frac{3+7\tilde{\mu}}{30}$ & $\frac{3+7\tilde{\mu}%
}{30}$\\
$\frac{3}{2}$ & $\frac{3}{2}$ & $\frac{3}{2}$ & $\frac{1+4\tilde{\mu}}{30}$ &
$\frac{1+4\tilde{\mu}}{30}$ & $\frac{1+4\tilde{\mu}}{30}$ & $\frac
{1+4\tilde{\mu}}{30}$ & $\frac{1+4\tilde{\mu}}{30}$ & $\frac{1+4\tilde{\mu}%
}{30}$\\
$\frac{5}{2}$ & $\frac{3}{2}$ & $\frac{1}{2}$ & $\times$ & $\times$ &
$\frac{-5-4\tilde{\mu}}{90}$ & $\frac{-5-4\tilde{\mu}}{90}$ & $\frac
{-5-4\tilde{\mu}}{90}$ & $\times$\\
$\frac{5}{2}$ & $\frac{3}{2}$ & $\frac{3}{2}$ & $\frac{5+17\tilde{\mu}}{90}$ &
$\frac{5+17\tilde{\mu}}{90}$ & $\frac{-1+2\tilde{\mu}}{90}$ & $\frac
{-1+29\tilde{\mu}}{630}$ & $\frac{-1+29\tilde{\mu}}{630}$ & $\frac
{41+134\tilde{\mu}}{630}$\\
$\frac{5}{2}$ & $\frac{5}{2}$ & $\frac{1}{2}$ & $\frac{1+2\tilde{\mu}}{6}$ &
$\frac{13+38\tilde{\mu}}{270}$ & $\frac{13+38\tilde{\mu}}{270}$ &
$\frac{-23+2\tilde{\mu}}{630}$ & $\frac{31+74\tilde{\mu}}{378}$ &
$\frac{31+74\tilde{\mu}}{378}$\\
$\frac{5}{2}$ & $\frac{5}{2}$ & $\frac{3}{2}$ & $\frac{29+104\tilde{\mu}}%
{630}$ & $\frac{23+152\tilde{\mu}}{1890}$ & $\frac{23+152\tilde{\mu}}{1890}$ &
$\frac{-1+8\tilde{\mu}}{210}$ & $\frac{55+232\tilde{\mu}}{1890}$ &
$\frac{55+232\tilde{\mu}}{1890}$\\
$\frac{5}{2}$ & $\frac{5}{2}$ & $\frac{5}{2}$ & $\frac{1+6\tilde{\mu}}{70}$ &
$\frac{1+6\tilde{\mu}}{70}$ & $\frac{1+6\tilde{\mu}}{70}$ & $\frac
{1+6\tilde{\mu}}{70}$ & $\frac{1+6\tilde{\mu}}{70}$ & $\frac{1+6\tilde{\mu}%
}{70}$\\
$\frac{7}{2}$ & $\frac{3}{2}$ & $\frac{3}{2}$ & $\times$ & $\times$ &
$\frac{-9-16\tilde{\mu}}{210}$ & $\frac{-9-16\tilde{\mu}}{210}$ &
$\frac{-9-16\tilde{\mu}}{210}$ & $\times$\\
$\frac{7}{2}$ & $\frac{5}{2}$ & $\frac{1}{2}$ & $\times$ & $\times$ &
$\frac{-7-8\tilde{\mu}}{126}$ & $\frac{-7-8\tilde{\mu}}{126}$ & $\frac
{-7-8\tilde{\mu}}{126}$ & $\times$\\
$\frac{7}{2}$ & $\frac{5}{2}$ & $\frac{3}{2}$ & $\frac{7+25\tilde{\mu}}{126}$
& $\frac{25+102\tilde{\mu}}{630}$ & $\frac{-20-11\tilde{\mu}}{1260}$ &
$\frac{-35-19\tilde{\mu}}{1890}$ & $\frac{-1+10\tilde{\mu}}{378}$ &
$\frac{40+149\tilde{\mu}}{756}$\\
$\frac{7}{2}$ & $\frac{5}{2}$ & $\frac{5}{2}$ & $\frac{133+668\tilde{\mu}%
}{5670}$ & $\frac{133+668\tilde{\mu}}{5670}$ & $\frac{-1+\tilde{\mu}}{210}$ &
$\frac{1+44\tilde{\mu}}{1134}$ & $\frac{1+44\tilde{\mu}}{1134}$ &
$\frac{11+52\tilde{\mu}}{378}$\\
$\frac{7}{2}$ & $\frac{7}{2}$ & $\frac{1}{2}$ & $\frac{1+2\tilde{\mu}}{6}$ &
$\frac{4+11\tilde{\mu}}{84}$ & $\frac{4+11\tilde{\mu}}{84}$ & $\frac
{-17-10\tilde{\mu}}{378}$ & $\frac{8+19\tilde{\mu}}{108}$ & $\frac
{8+19\tilde{\mu}}{108}$\\
$\frac{7}{2}$ & $\frac{7}{2}$ & $\frac{3}{2}$ & $\frac{19+68\tilde{\mu}}{378}$
& $\frac{1+8\tilde{\mu}}{126}$ & $\frac{1+8\tilde{\mu}}{126}$ & $\frac
{-11-4\tilde{\mu}}{630}$ & $\frac{47+208\tilde{\mu}}{1890}$ & $\frac
{47+208\tilde{\mu}}{1890}$\\
$\frac{7}{2}$ & $\frac{7}{2}$ & $\frac{5}{2}$ & $\frac{43+218\tilde{\mu}%
}{1890}$ & $\frac{4+41\tilde{\mu}}{756}$ & $\frac{4+41\tilde{\mu}}{756}$ &
$\frac{-1+10\tilde{\mu}}{378}$ & $\frac{56+331\tilde{\mu}}{3780}$ &
$\frac{56+331\tilde{\mu}}{3780}$\\
$\frac{7}{2}$ & $\frac{7}{2}$ & $\frac{7}{2}$ & $\frac{1+8\tilde{\mu}}{126}$ &
$\frac{1+8\tilde{\mu}}{126}$ & $\frac{1+8\tilde{\mu}}{126}$ & $\frac
{1+8\tilde{\mu}}{126}$ & $\frac{1+8\tilde{\mu}}{126}$ & $\frac{1+8\tilde{\mu}%
}{126}$%
\end{tabular}
\end{center}
\caption{Mean values $\left\langle \mathbb{S}_{z}\right\rangle _{c}$ of
three-fermion states $\left\vert (j_{1},j_{2},j_{3},J_{c})1/2,1/2\right\rangle
$ with $J_{c}=j_{c}-1/2$ and $J_{c}=j_{c}+1/2$ (columns 3,4,5 and 5,6,7;
$c=3,2,1$), see the first relation (\ref{rs13}) and (\ref{rs17}). The symbol
$\times$ denotes configuration for which the second condition (\ref{rs13}) is
not satisfied.}%
\label{tb2}%
\end{table}The meaning of the parameter $\tilde{\mu}$ is as follows:

\textit{i)} If one assumes the same parameter $\mu\ $(\ref{rs4a}) for the
three fermions in the state (\ref{rs12}), then $\tilde{\mu}=\mu.$

\textit{ii)} In a general case the complete wave function%
\begin{equation}
\Psi_{c,1/2,1/2}=\phi_{j_{1}}\left(  \epsilon_{1}\right)  \phi_{j_{2}}\left(
\epsilon_{2}\right)  \phi_{j_{3}}\left(  \epsilon_{3}\right)  \Phi
_{c,1/2,1/2}(\mathbf{\omega}_{1},\mathbf{\omega}_{2},\mathbf{\omega}_{3})
\label{rs18a}%
\end{equation}
gives instead of (\ref{rs8b}) a more complicated expression%
\begin{equation}
\tilde{\mu}=f_{c}\left(  \left\langle \mu_{1}\right\rangle ,\left\langle
\mu_{2}\right\rangle ,\left\langle \mu_{3}\right\rangle ,j_{1},j_{2}%
,j_{3}\right)  , \label{rs18b}%
\end{equation}
where the parameters $\left\langle \mu_{i}\right\rangle $\ are defined by Eq.
(\ref{rs8b}). The expression is simplified for $\left\langle \mu
_{1}\right\rangle =\left\langle \mu_{2}\right\rangle =\left\langle \mu
_{3}\right\rangle =\left\langle \mu\right\rangle $ :%
\begin{equation}
f_{c}\left(  \left\langle \mu\right\rangle ,\left\langle \mu\right\rangle
,\left\langle \mu\right\rangle ,j_{1},j_{2},j_{3}\right)  =\left\langle
\mu\right\rangle . \label{rs18c}%
\end{equation}
Just for illustration (\textit{obtained with WM}), the corresponding
$\tilde{\mu}$ in the third row and last column of the table reads:%
\begin{equation}
\tilde{\mu}=f_{1}\left(  \left\langle \mu_{1}\right\rangle ,\left\langle
\mu_{2}\right\rangle ,\left\langle \mu_{3}\right\rangle ,3/2,3/2,1/2\right)
=\frac{-4\left\langle \mu_{1}\right\rangle +6\left\langle \mu_{2}\right\rangle
+5\left\langle \mu_{3}\right\rangle }{7} \label{rs18d}%
\end{equation}
and in general
\begin{equation}
\tilde{\mu}=f_{c}\left(  \left\langle \mu_{1}\right\rangle ,\left\langle
\mu_{2}\right\rangle ,\left\langle \mu_{3}\right\rangle ,j_{1},j_{2}%
,j_{3}\right)  =\frac{n_{1}\left\langle \mu_{1}\right\rangle +n_{2}%
\left\langle \mu_{2}\right\rangle +n_{3}\left\langle \mu_{3}\right\rangle
}{n_{1}+n_{2}+n_{3}},\qquad\left\vert \tilde{\mu}\right\vert \leq1,
\label{rs18f}%
\end{equation}
where the $n_{i}$ depend on $j_{k}$. Obviously the many-fermion system with
$J=J_{z}=1/2$ can be treated as a composed particle of the spin $1/2$. This
spin is generated by the spins and OAMs of the involved fermions. The relative
weights of the spin and OAM contributions vary depending not only on the
intrinsic values $j_{1},j_{2},j_{3}$ and the pattern of composition, but also
on the mass-motion parameter $\tilde{\mu}$. The data in the table suggest that
for any configuration in the relativistic limit $\tilde{\mu}\rightarrow0$ we
have%
\begin{equation}
\left\vert \left\langle \mathbb{S}_{z}\right\rangle \right\vert \leq\frac
{1}{6} \label{rs19}%
\end{equation}
similarly as in the case of the one-fermion states (\ref{rs8}).

The table illustrates a complexity of the AM composition even for only three
fermions. Is there a simple rule like (\ref{rs19}) for $n>3$? First, let us
consider the composition%
\begin{equation}
\Psi_{c,1/2,1/2}=\left\vert (j_{1},j_{2},...j_{n})_{c}1/2,1/2\right\rangle ,
\label{rs21}%
\end{equation}
where all one-fermion AMs are the same, $j_{i\text{ }}=j$ (like the rows
1,4,9,17 in table). The corresponding spin reads%
\begin{equation}
\left\langle \mathbb{S}_{z}\right\rangle =\frac{1+\left(  2j+1\right)
\tilde{\mu}}{8j\left(  j+1\right)  } \label{RS23}%
\end{equation}
regardless of $n$ and details of composition. The proof of this relation is
given in Appendix \ref{app1}. Apparently for $\tilde{\mu}\rightarrow
0$\ the\ relation (\ref{rs19}) is again satisfied. The situation with the
composition of different AMs is getting much more complex for increasing $n$.
However, an average value of the spin over all possible composition patterns
of the state $\left\vert (j_{1},j_{2},...j_{n})_{c}1/2,1/2\right\rangle $
appear (\textit{obtained with WM)} to safely satisfy (\ref{rs19}). This is the
case when there is no (e.g. dynamical) preference among various composition patterns.

Let us illustrate a possible role of the composition patterns by the simple
example $j_{1},j_{2},j_{3}=1/2$. Eq. (\ref{rs12}) gives the three\ states
corresponding to $J_{c}=1$:%
\begin{equation}
\Psi_{abc,1/2,1/2}=\frac{\phi_{abc}}{\sqrt{6}}\left(  \left\vert
-1/2,1/2,1/2\right\rangle +\left\vert 1/2,-1/2,1/2\right\rangle -2\left\vert
1/2,1/2,-1/2\right\rangle \right)  , \label{ri2}%
\end{equation}
where%
\begin{equation}
\phi_{abc}=\phi_{a}\left(  \epsilon_{a}\right)  \phi_{b}\left(  \epsilon
_{b}\right)  \phi_{c}\left(  \epsilon_{c}\right)  . \label{ri3}%
\end{equation}
The indices $abc$ define the composition in accordance with (\ref{ri1}), AM
states are defined correspondingly: $\left\vert j_{za},j_{zb},j_{zc}%
\right\rangle $. The other three states correspond to $J_{c}=0$:
\begin{equation}
\Psi_{abc,1/2,1/2}=\frac{\phi_{abc}}{\sqrt{2}}\left(  \left\vert
1/2,-1/2,1/2\right\rangle -\left\vert -1/2,1/2,1/2\right\rangle \right)  .
\label{ri4}%
\end{equation}
Now let us remind the non-relativistic proton SU(6) wave function in the
standard notation:%
\begin{equation}
\left\vert p\uparrow\right\rangle =\frac{1}{\sqrt{2}}\left\{  \frac{1}%
{\sqrt{6}}\left\vert duu+udu-2uud\right\rangle \frac{1}{\sqrt{6}}\left\vert
\downarrow\uparrow\uparrow+\uparrow\downarrow\uparrow-2\uparrow\uparrow
\downarrow\right\rangle +\frac{1}{\sqrt{2}}\left\vert duu-udu\right\rangle
\frac{1}{\sqrt{2}}\left\vert \downarrow\uparrow\uparrow-\uparrow
\downarrow\uparrow\right\rangle \right\}  . \label{ri5}%
\end{equation}
The comparison (\ref{ri2})-(\ref{ri4}) with (\ref{ri5}) suggest the SU(6) wave
function after substitution%
\[
\phi_{a}\left(  \epsilon_{a}\right)  =u_{1},\qquad\phi_{b}\left(  \epsilon
_{b}\right)  =u_{2,}\qquad\phi_{c}\left(  \epsilon_{c}\right)  =d
\]
can be obtained as the superposition of wave functions generated by the AM
compositions%
\begin{equation}
(\left(  u_{1}\oplus u_{2}\right)  _{J}\oplus d)_{1/2},\qquad(\left(  d\oplus
u_{1}\right)  _{J}\oplus u_{2})_{1/2},\qquad(\left(  u_{2}\oplus d\right)
_{J}\oplus u_{1})_{1/2} \label{ri6}%
\end{equation}
for $J=1,2$.

\section{Distributions and structure functions}

\label{D&SFs}In this section we will study the distribution and structure
functions (SFs) of the quark states, which are represented by the free fermion
eigenstates of angular momentum as described in the previous section. A
particular case of these functions has been discussed in our previous study
\cite{Zavada:2007ww} that was focused on the quark state $j=$ $j_{z}%
=1/2,\ l_{p}=0$. This case is equivalent to the state (\ref{rs12}) with
$j_{1}=$ $j_{2}=$ $j_{3}=1/2$ and $l_{p1}=$ $l_{p2}=$ $l_{p3}=0$. But the
situation with the SFs, which are generated by the states with $j_{k}>1/2$ or
$l_{pk}>0$ is more intricate.

\subsection{Polarized distributions and spin vectors}

First, we define the projectors%
\begin{equation}
\mathcal{P}_{\lambda,\pm}=\left(
\begin{array}
[c]{cc}%
\sigma_{\lambda,\pm} & 0\\
0 & \frac{\mathbf{p\sigma}}{\epsilon+m}\sigma_{\lambda,\pm}\frac
{\mathbf{p\sigma}}{\epsilon-m}%
\end{array}
\right)  , \label{rs31}%
\end{equation}
where%
\begin{equation}
\sigma_{\lambda,\pm}=\frac{1}{2}\left(  \mathbf{1\pm}\sigma_{\lambda}\right)
\label{rs32}%
\end{equation}
and $\mathbf{\sigma}=(\sigma_{x},\sigma_{y},\sigma_{z})$ are Pauli matrices.
Obviously
\begin{gather}
\mathcal{P}_{\lambda,+}+\mathcal{P}_{\lambda,-}=\mathbf{1},\qquad
\mathcal{P}_{\lambda,+}\mathcal{P}_{\lambda,-}=\mathcal{P}_{\lambda
,-}\mathcal{P}_{\lambda,+}=\mathbf{0},\qquad\left(  \mathcal{P}_{\lambda,\pm
}\right)  ^{2}=\mathcal{P}_{\lambda,\pm},\label{rs33}\\
\Delta\mathcal{P}_{\lambda}\equiv\mathcal{P}_{\lambda,+}-\mathcal{P}%
_{\lambda,-}=\left(
\begin{array}
[c]{cc}%
\sigma_{\lambda} & 0\\
0 & \frac{\mathbf{p\sigma}}{\epsilon+m}\sigma_{\lambda}\frac{\mathbf{p\sigma}%
}{\epsilon-m}%
\end{array}
\right)  . \label{rs34}%
\end{gather}
Apparently any solution of Dirac equation (including the states (\ref{rs1}))
\begin{equation}
\Phi=\left(
\begin{array}
[c]{c}%
\varphi\\
\frac{\mathbf{p\sigma}}{\epsilon+m}\varphi
\end{array}
\right)  \label{rs35}%
\end{equation}
can be expressed as a superposition%
\begin{equation}
\Phi=\Phi_{\lambda,+}+\Phi_{\lambda,-},\qquad\Phi_{\lambda,\pm}=\mathcal{P}%
_{\lambda,\pm}\Phi, \label{rs36}%
\end{equation}
where $\Phi_{\lambda,\pm}$ are states with positive or negative polarization
in direction of axis $\lambda$ of the quark \textit{rest frame}. The states
(\ref{rs1}) generate polarization distributions%
\begin{equation}
H_{\lambda,j,l_{p},j_{z}}\left(  \mathbf{\omega}\right)  =\Phi_{jl_{p}j_{z}%
}^{+}\Delta\mathcal{P}_{\lambda}\Phi_{jl_{p}j_{z}}. \label{rs38}%
\end{equation}
Some examples of this distribution for $\lambda=z$ are given in Tab.
\ref{tb1}. One can calculate the integrals (\textit{obtained with WM}):
\begin{equation}
\int H_{z,j,j\mp1/2,j_{z}}\left(  \mathbf{\omega}\right)  d\mathbf{\omega}=%
\genfrac{\{}{\}}{0pt}{}{\frac{j_{z}}{j}}{\frac{-j_{z}}{j+1}}%
. \label{rs38b}%
\end{equation}
Let us note, the last relation and the first relation (\ref{rs6}) coincide,
since the both are equivalent definitions of the spin projections in the
\ non-relativistic limit:%
\begin{equation}
\left\langle s_{z}\right\rangle _{NR}=\frac{1}{2}\int H_{j,j\mp1/2,j_{z}%
}\left(  \mathbf{\omega}\right)  d\mathbf{\omega.} \label{rs38a}%
\end{equation}

In the next step we will discuss polarized distributions related to the
many-quark states. The polarized counterpart to\ the average distribution
(\ref{rs11a}) reads:%
\begin{equation}
h_{\lambda,c,k}(\mathbf{\omega}_{k})=\int\Phi_{c,1/2,1/2}^{+}\Delta
\mathcal{P}_{\lambda,k}\Phi_{c,1/2,1/2}%
{\displaystyle\prod\limits_{i\neq k}^{n}}
d\mathbf{\omega}_{i}. \label{RS39}%
\end{equation}
A particular case of this distribution is discussed in Appendix \ref{app1}. In
general this distribution does not have rotational symmetry like the
corresponding unpolarized distribution (\ref{rs11a}), but has the form
(\textit{obtained with WM}):
\begin{gather}
h_{x,c,k}(\mathbf{\omega})=\frac{1}{4\pi}\beta_{c,k}\sin2\theta\cos
\varphi,\qquad h_{y,c,k}(\mathbf{\omega})=\frac{1}{4\pi}\beta_{c,k}\sin
2\theta\sin\varphi,\label{rs41}\\
h_{z,c,k}(\mathbf{\omega})=\frac{1}{4\pi}\left(  \alpha_{c,k}+\beta_{c,k}%
\cos2\theta\right)  , \label{rs40}%
\end{gather}
where the constants $\alpha_{c,k}$ and $\beta_{c,k}$ depend on the pattern of
composition and absorb corresponding Clebsch-Gordan coefficients entering
matrix elements (\ref{RS39}).

Let us consider for an illustration the three-quark states (\ref{rs12}). The
corresponding total one-quark polarized distributions read%
\begin{equation}
H_{\lambda,c}\left(  \mathbf{\omega}\right)  =\sum_{k=1}^{3}h_{\lambda
,c,k}\left(  \mathbf{\omega}\right)  , \label{rs43}%
\end{equation}
where $h_{\lambda,c,k}$\ are defined by (\ref{RS39}) for $n=3.$ The form of
resulting distributions $H_{\lambda,c}$ follow from (\ref{rs41}) $-$
(\ref{rs43}),%
\begin{gather}
H_{x,c}\left(  \mathbf{\omega}\right)  =b_{c}\sin2\theta\cos\varphi
,\label{rs44a}\\
H_{y,c}\left(  \mathbf{\omega}\right)  =b_{c}\sin2\theta\sin\varphi
,\label{rs44b}\\
H_{z,c}\left(  \mathbf{\omega}\right)  =a_{c}+b_{c}\cos2\theta, \label{rs44c}%
\end{gather}
where%
\begin{equation}
a_{c}=\frac{1}{4\pi}\sum_{k=1}^{3}\alpha_{c,k},\qquad b_{c}=\frac{1}{4\pi}%
\sum_{k=1}^{3}\beta_{c,k}. \label{rs44d}%
\end{equation}
The distributions $H_{z,c}$ with the factors $a_{c},b_{c}$ are given in Tab.
\ref{tb3} for a set of input values $j_{1},j_{2},j_{3}$ (\textit{obtained with
WM}). The table is displayed for $l_{p}=j-1/2$, but for opposite choice
$l_{p}=j+1/2$ both factors are simply interchanged: $b_{c}\rightleftarrows
a_{c}$, like in Tab. \ref{ta1}. We have verified that if we calculate the
unpolarized distributions $P_{c}(\mathbf{\omega})$ instead of the polarized
$H_{z,c}$, then in an agreement with Eq. (\ref{rs16b}) we get $3/4\pi$ in any
position of the table. \begin{table}[ptb]
\begin{center}%
\begin{tabular}
[c]{ccc|ccc|ccc}%
$j_{1}$ & $j_{2}$ & $j_{3}$ & $H_{3}$ & $H_{2}$ & $H_{1}$ & $H_{3}$ & $H_{2}$
& $H_{1}$\\\hline
$\frac{1}{2}$ & $\frac{1}{2}$ & $\frac{1}{2}$ & $1$ & $1$ & $1$ & $1$ & $1$ &
$1$\\
$\frac{3}{2}$ & $\frac{1}{2}$ & $\frac{1}{2}$ & $\times$ & $\times$ &
$\frac{-1-\cos2\theta}{6}$ & $\frac{-1-\cos2\theta}{6}$ & $\frac
{-1-\cos2\theta}{6}$ & $\times$\\
$\frac{3}{2}$ & $\frac{3}{2}$ & $\frac{1}{2}$ & $1$ & $\frac{5-\cos2\theta
}{12}$ & $\frac{5-\cos2\theta}{12}$ & $\frac{1-2\cos2\theta}{15}$ &
$\frac{13-\cos2\theta}{20}$ & $\frac{13-\cos2\theta}{20}$\\
$\frac{3}{2}$ & $\frac{3}{2}$ & $\frac{3}{2}$ & $\frac{3-\cos2\theta}{10}$ &
$\frac{3-\cos2\theta}{10}$ & $\frac{3-\cos2\theta}{10}$ & $\frac{3-\cos
2\theta}{10}$ & $\frac{3-\cos2\theta}{10}$ & $\frac{3-\cos2\theta}{10}$\\
$\frac{5}{2}$ & $\frac{3}{2}$ & $\frac{1}{2}$ & $\times$ & $\times$ &
$\frac{-7-3\cos2\theta}{30}$ & $\frac{-7-3\cos2\theta}{30}$ & $\frac
{-7-3\cos2\theta}{30}$ & $\times$\\
$\frac{5}{2}$ & $\frac{3}{2}$ & $\frac{3}{2}$ & $\frac{27-7\cos2\theta}{60}$ &
$\frac{27-7\cos2\theta}{60}$ & $\frac{-\cos2\theta}{15}$ & $\frac
{27-31\cos2\theta}{420}$ & $\frac{27-31\cos2\theta}{420}$ & $\frac
{54-13\cos2\theta}{105}$\\
$\frac{5}{2}$ & $\frac{5}{2}$ & $\frac{1}{2}$ & $1$ & $\frac{16-3\cos2\theta
}{45}$ & $\frac{16-3\cos2\theta}{45}$ & $\frac{-11-12\cos2\theta}{105}$ &
$\frac{34-3\cos2\theta}{63}$ & $\frac{34-3\cos2\theta}{63}$\\
$\frac{5}{2}$ & $\frac{5}{2}$ & $\frac{3}{2}$ & $\frac{81-23\cos2\theta}{210}$
& $\frac{99-53\cos2\theta}{630}$ & $\frac{99-53\cos2\theta}{630}$ &
$\frac{3-5\cos2\theta}{70}$ & $\frac{171-61\cos2\theta}{630}$ & $\frac
{171-61\cos2\theta}{630}$\\
$\frac{5}{2}$ & $\frac{5}{2}$ & $\frac{5}{2}$ & $\frac{6-3\cos2\theta}{35}$ &
$\frac{6-3\cos2\theta}{35}$ & $\frac{6-3\cos2\theta}{35}$ & $\frac
{6-3\cos2\theta}{35}$ & $\frac{6-3\cos2\theta}{35}$ & $\frac{6-3\cos2\theta
}{35}$\\
$\frac{7}{2}$ & $\frac{3}{2}$ & $\frac{3}{2}$ & $\times$ & $\times$ &
$\frac{-17-\cos2\theta}{70}$ & $\frac{-17-\cos2\theta}{70}$ & $\frac
{-17-\cos2\theta}{70}$ & $\times$\\
$\frac{7}{2}$ & $\frac{5}{2}$ & $\frac{1}{2}$ & $\times$ & $\times$ &
$\frac{-11-3\cos2\theta}{42}$ & $\frac{-11-3\cos2\theta}{42}$ & $\frac
{-11-3\cos2\theta}{42}$ & $\times$\\
$\frac{7}{2}$ & $\frac{5}{2}$ & $\frac{3}{2}$ & $\frac{39-11\cos2\theta}{84}$
& $\frac{38-13\cos2\theta}{105}$ & $\frac{-51-29\cos2\theta}{840}$ &
$\frac{-89-51\cos2\theta}{1260}$ & $\frac{2-3\cos2\theta}{63}$ &
$\frac{229-69\cos2\theta}{504}$\\
$\frac{7}{2}$ & $\frac{5}{2}$ & $\frac{5}{2}$ & $\frac{467-201\cos2\theta
}{1890}$ & $\frac{467-201\cos2\theta}{1890}$ & $\frac{1-3\cos2\theta}{70}$ &
$\frac{23-21\cos2\theta}{378}$ & $\frac{23-21\cos2\theta}{378}$ &
$\frac{37-15\cos2\theta}{126}$\\
$\frac{7}{2}$ & $\frac{7}{2}$ & $\frac{1}{2}$ & $1$ & $\frac{19-3\cos2\theta
}{56}$ & $\frac{19-3\cos2\theta}{56}$ & $\frac{-11-6\cos2\theta}{63}$ &
$\frac{35-3\cos2\theta}{72}$ & $\frac{35-3\cos2\theta}{72}$\\
$\frac{7}{2}$ & $\frac{7}{2}$ & $\frac{3}{2}$ & $\frac{53-15\cos2\theta}{56}$
& $\frac{5-3\cos2\theta}{42}$ & $\frac{5-3\cos2\theta}{42}$ & $\frac
{-13-9\cos2\theta}{210}$ & $\frac{151-57\cos2\theta}{630}$ & $\frac
{151-57\cos2\theta}{630}$\\
$\frac{7}{2}$ & $\frac{7}{2}$ & $\frac{5}{2}$ & $\frac{76-33\cos2\theta}{315}$
& $\frac{49-33\cos2\theta}{504}$ & $\frac{49-33\cos2\theta}{504}$ &
$\frac{2-3\cos2\theta}{63}$ & $\frac{443-219\cos2\theta}{2520}$ &
$\frac{443-219\cos2\theta}{2520}$\\
$\frac{7}{2}$ & $\frac{7}{2}$ & $\frac{7}{2}$ & $\frac{5-3\cos2\theta}{42}$ &
$\frac{5-3\cos2\theta}{42}$ & $\frac{5-3\cos2\theta}{42}$ & $\frac
{5-3\cos2\theta}{42}$ & $\frac{5-3\cos2\theta}{42}$ & $\frac{5-3\cos2\theta
}{42}$%
\end{tabular}
\end{center}
\caption{Polarized distributions $H_{z,c}$ (common factor $1/4\pi$ is omitted)
generated by the three-fermion states $\left\vert (j_{1},j_{2},j_{3}%
,J_{c})1/2,1/2\right\rangle $ with $J_{c}=j_{c}-1/2$ and $J_{c}=j_{c}+1/2$
(columns 3,4,5 and 5,6,7; $c=3,2,1$), see the first relation (\ref{rs13}) and
relation (\ref{rs44c}). The symbol $\times$ denotes configuration for which
the second condition (\ref{rs13}) is not satisfied.}%
\label{tb3}%
\end{table}Let us note the correspondence between the tables \ref{tb3} and
\ref{tb2} for $l_{p}=j\mp1/2$ and $\mu=\pm1,$
\begin{equation}
\frac{1}{2}\int H_{z,c}\left(  \mathbf{\omega}\right)  d\mathbf{\omega
=}\left\langle \mathbb{S}_{z}\right\rangle _{c,NR}, \label{rs45}%
\end{equation}
which agree with Eq. (\ref{rs38a}). At the same time we have%
\begin{equation}
\frac{1}{2}\int H_{x,c}\left(  \mathbf{p}\right)  d^{3}\mathbf{p=}\left\langle
\mathbb{S}_{z}\right\rangle _{c,NR}=0,\qquad\frac{1}{2}\int H_{y,c}\left(
\mathbf{p}\right)  d^{3}\mathbf{p=}\left\langle \mathbb{S}_{z}\right\rangle
_{c,NR}=0. \label{rs46a}%
\end{equation}
Apparently, the distributions (\ref{rs44a})$-$(\ref{rs44c}) are representation
of the quark spin vector $\mathbf{w}\left(  \mathbf{\omega}\right)  $ in
spherical coordinates%
\begin{equation}
(H_{z}\left(  \mathbf{\omega}\right)  ,H_{x}\left(  \mathbf{\omega}\right)
,H_{y}\left(  \mathbf{\omega}\right)  )=\mathbf{w}\left(  \mathbf{\omega
}\right)  \label{rs45a}%
\end{equation}
in the non-relativistic limit (rearrangement of axes is just for convenience).
The vector $\mathbf{w}$\ can be modified as%
\begin{equation}
\mathbf{w}\left(  \mathbf{\omega}\right)  =\left(  a_{c}-b_{c}+2b_{c}\cos
^{2}\theta,2b_{c}\cos\theta\sin\theta\cos\varphi,2b_{c}\cos\theta\sin
\theta\sin\varphi\right)  , \label{rs45c}%
\end{equation}
which can be represented as%
\begin{equation}
\mathbf{w}\left(  \mathbf{\omega}\right)  =\left(  a_{c}-b_{c}\right)
\mathbf{S}+2b_{c}\left(  \mathbf{n\cdot S}\right)  \mathbf{n,} \label{rs45b}%
\end{equation}
where $\mathbf{n=p/}\left\vert \mathbf{p}\right\vert $ and $\mathbf{S}$ is the
unit vector defining the axis of $j_{z}$ projections, which is identical to
the proton spin vector in the proton rest frame. If we replace (\ref{rs12}) by
the complete wave function (\ref{rs18a}), then the factors (\ref{rs44d}) are
replaced by%
\begin{equation}
\mathrm{u}\left(  \epsilon\right)  =\sum_{k=1}^{3}\alpha_{c,k}a_{j_{k}}^{\ast
}\left(  \epsilon\right)  a_{j_{k}}\left(  \epsilon\right)  ,\qquad
\mathrm{v}\left(  \epsilon\right)  =\sum_{k=1}^{3}\beta_{c,k}a_{j_{k}}^{\ast
}\left(  \epsilon\right)  a_{j_{k}}\left(  \epsilon\right)  , \label{rs47d}%
\end{equation}
and the vector (\ref{rs45b}) is modified correspondingly:%
\begin{equation}
\mathbf{w}\left(  \mathbf{\omega,}\epsilon\right)  =\left(  \mathrm{u}\left(
\epsilon\right)  -\mathrm{v}\left(  \epsilon\right)  \right)  \mathbf{S}%
+2\mathrm{v}\left(  \epsilon\right)  \left(  \mathbf{n\cdot S}\right)
\mathbf{n.} \label{rs47c}%
\end{equation}
It is convenient to define the constants $\mathrm{U,V}$:
\begin{equation}
\mathrm{U}\equiv\int\mathrm{u}\left(  \epsilon\right)  d^{3}\mathbf{p=}%
\sum_{k=1}^{3}\alpha_{c,k},\qquad\mathrm{V}\equiv\int\mathrm{v}\left(
\epsilon\right)  d^{3}\mathbf{p=}\sum_{k=1}^{3}\beta_{c,k}. \label{rs46b}%
\end{equation}
The relations (\ref{rs47d}) define the scalar functions depending on the
parameter $\epsilon=p\cdot P/M$, which is the quark energy in the nucleon rest
frame. The form of the spin vector (\ref{rs47c}) is characteristic for
\textit{any system} $J=1/2$ \textit{regardless of the number of involved
quarks}. In the first row of Tab. \ref{tb1} we have two possibilities for
$H_{z}$ corresponding to $j=j_{z}=1/2$, so in general this distribution will
be the combination%
\begin{equation}
H_{z}\left(  \mathbf{\omega,}\epsilon\right)  =\mathrm{u}\left(
\epsilon\right)  +\mathrm{v}\left(  \epsilon\right)  \cos2\theta\label{rs48}%
\end{equation}
and correspondingly for $H_{x},H_{y}$%
\begin{equation}
H_{x}\left(  \mathbf{\omega,}\epsilon\right)  =\mathrm{v}\left(
\epsilon\right)  \sin2\theta\cos\varphi,\qquad H_{y}\left(  \mathbf{\omega
,}\epsilon\right)  =\mathrm{v}\left(  \epsilon\right)  \sin2\theta\sin\varphi,
\label{rs48a}%
\end{equation}
which can be equivalently represented by (\ref{rs47c}).

\subsection{Spin structure functions}

The spin SFs can be extracted from the antisymmetric part of hadronic tensor
in a similar way as done in \cite{Zavada:2001bq}. General form of this tensor
reads%
\begin{equation}
T_{\alpha\beta}^{(A)}=\varepsilon_{\alpha\beta\lambda\sigma}q^{\lambda}\left(
MS^{\sigma}G_{1}+\left(  (P\cdot q)S^{\sigma}-(q\cdot S)P^{\sigma}\right)
\frac{G_{2}}{M}\right)  , \label{cr18}%
\end{equation}
which after substitution%
\begin{equation}
G_{S}=MG_{1}+\frac{P\cdot q}{M}G_{2},\quad G_{P}=\frac{q\cdot S}{M}G_{2},
\label{cr19}%
\end{equation}
gives%
\begin{equation}
T_{\alpha\beta}^{(A)}=\varepsilon_{\alpha\beta\lambda\sigma}q^{\lambda}\left(
S^{\sigma}G_{S}-P^{\sigma}G_{P}\right)  . \label{cr20}%
\end{equation}
The spin SFs in the standard notation $g_{1}=\left(  P\cdot q\right)  MG_{1},$
$g_{2}=\left(  \left(  P\cdot q\right)  ^{2}/M\right)  G_{2}$ satisfy%
\begin{equation}
g_{1}=\left(  P\cdot q\right)  \left(  G_{S}-\frac{P\cdot q}{q\cdot S}%
G_{P}\right)  ,\qquad g_{2}=\frac{\left(  P\cdot q\right)  ^{2}}{q\cdot
S}G_{P},\qquad g_{1}+g_{2}=\left(  P\cdot q\right)  G_{S}. \label{cra31}%
\end{equation}
In the next, to simplify the related expressions, if not stated otherwise we
ignore different quark flavors and consider the quark charges equal unity. The
antisymmetric part of the tensor related to a plane wave with momentum $p$
reads
\begin{equation}
t_{\alpha\beta}^{(A)}=m\varepsilon_{\alpha\beta\lambda\sigma}q^{\lambda
}w^{\sigma}(p) \label{cs5}%
\end{equation}
so the full tensor is given by the integral:%
\begin{equation}
T_{\alpha\beta}^{(A)}=\varepsilon_{\alpha\beta\lambda\sigma}q^{\lambda}m\int
w^{\sigma}(p)\delta((p+q)^{2}-m^{2})\frac{d^{3}\mathbf{p}}{\epsilon}.
\label{cr5}%
\end{equation}
The quark spin vector $w$ can be written in the manifestly covariant form%
\begin{equation}
w^{\sigma}=AP^{\sigma}+BS^{\sigma}+Cp^{\sigma}, \label{cr15}%
\end{equation}
where $A,B,C$ are invariant functions (scalars) of the relevant vectors
$P,S,p$ \cite{Zavada:2001bq}. These three functions are fixed by the condition
$pw=0$ and by the form of the spin vector in the quark rest frame
(\ref{rs47c}). In the Appendix \ref{app2} we have proved:%
\begin{align}
A  &  =-\left(  p\cdot S\right)  \left(  \frac{\mathrm{u}\left(
\epsilon\right)  }{p\cdot P+mM}-\frac{\mathrm{v}\left(  \epsilon\right)
}{p\cdot P-mM}\right)  ,\label{rs61}\\
B  &  =\mathrm{u}\left(  \epsilon\right)  -\mathrm{v}\left(  \epsilon\right)
,\label{rs62}\\
C  &  =-\left(  p\cdot S\right)  \frac{M}{m}\left(  \frac{\mathrm{u}\left(
\epsilon\right)  }{p\cdot P+mM}+\frac{\mathrm{v}\left(  \epsilon\right)
}{p\cdot P-mM}\right)  . \label{rs63}%
\end{align}
The comparison of Eqs. (\ref{cr20}) and (\ref{cr5}) gives%
\begin{equation}
\varepsilon_{\alpha\beta\lambda\sigma}q^{\lambda}\left(  S^{\sigma}%
G_{S}-P^{\sigma}G_{P}\right)  =\varepsilon_{\alpha\beta\lambda\sigma
}q^{\lambda}\frac{m}{2P\cdot q}\int w^{\sigma}(p)\delta\left(  \frac{p\cdot
q}{P\cdot q}-x\right)  \frac{d^{3}\mathbf{p}}{\epsilon}, \label{rs64}%
\end{equation}
where we have modified the $\delta-$function term%
\begin{equation}
\delta((p+q)^{2}-m^{2})=\frac{1}{2P\cdot q}\delta\left(  \frac{p\cdot
q}{P\cdot q}-x\right)  \label{rs65}%
\end{equation}
with the Bjorken variable $x=Q^{2}/\left(  2P\cdot q\right)  $. Because of
antisymmetry of the tensor $\varepsilon$ it follows that
\begin{equation}
S^{\sigma}G_{S}-P^{\sigma}G_{P}=\frac{m}{2P\cdot q}\int w^{\sigma}%
(p)\delta\left(  \frac{p\cdot q}{P\cdot q}-x\right)  \frac{d^{3}\mathbf{p}%
}{\epsilon}+Dq^{\sigma}, \label{cr25}%
\end{equation}
where $D$ is a scalar function. After contracting with $P_{\sigma},S_{\sigma}$
and $q_{\sigma}$ (and taking into account $P^{2}=M,\quad PS=0,\quad S^{2}=-1$)
one gets the equations for unknown functions $G_{S},G_{P}$ and $D$:%
\begin{align}
-M^{2}G_{P}  &  =\left\{  P\cdot w\right\}  +D\left(  P\cdot q\right)
,\label{cr26}\\
-G_{S}  &  =\left\{  S\cdot w\right\}  +D\left(  q\cdot S\right)
,\label{cr27}\\
\left(  q\cdot S\right)  G_{S}-\left(  P\cdot q\right)  G_{P}  &  =\left\{
q\cdot w\right\}  +Dq^{2}, \label{cr28}%
\end{align}
where we used the compact notation:%
\begin{equation}
\left\{  yy\right\}  \equiv\frac{m}{2P\cdot q}\int\left(  yy\right)
\delta\left(  \frac{p\cdot q}{P\cdot q}-x\right)  \frac{d^{3}\mathbf{p}%
}{\epsilon}. \label{cr29}%
\end{equation}
The function $D$ can be easily extracted%
\begin{equation}
D=\frac{\left\{  P\cdot w\right\}  \left(  P\cdot q\right)  /M^{2}-\left\{
S\cdot w\right\}  \left(  q\cdot S\right)  -\left\{  q\cdot w\right\}  }%
{q^{2}+\left(  q\cdot S\right)  ^{2}-\left(  P\cdot q/M\right)  ^{2}}.
\label{cr30}%
\end{equation}
The explicit form of expressions $\left\{  X\cdot w\right\}  $ follows from
Eqs. (\ref{cr15})$-$(\ref{rs63})%
\begin{align}
P\cdot w  &  =AM+C\left(  p\cdot P\right)  ,\label{cr31}\\
S\cdot w  &  =-B+C\left(  p\cdot S\right)  ,\label{cr32}\\
q\cdot w  &  =A\left(  P\cdot q\right)  +B\left(  S\cdot q\right)  +C\left(
p\cdot S\right)  , \label{cr33}%
\end{align}
which after substitution to Eqs. (\ref{cr26}),(\ref{cr27}) and (\ref{cr30})
gives the functions $G_{P}$ and $G_{S}$. The details of this calculation are
explained in the Appendix \ref{app3}, where we obtained the relations%
\begin{align}
g_{1}\left(  x\right)   &  =\frac{1}{2}\int\left(  \mathrm{u}\left(
\epsilon\right)  \left(  p_{1}+m+\frac{p_{1}^{2}}{\epsilon+m}\right)
+\mathrm{v}\left(  \epsilon\right)  \left(  p_{1}-m+\frac{p_{1}^{2}}%
{\epsilon-m}\right)  \right)  \delta\left(  \frac{\epsilon+p_{1}}{M}-x\right)
\frac{d^{3}\mathbf{p}}{\epsilon},\label{cr34}\\
g_{2}\left(  x\right)   &  =-\frac{1}{2}\int\left(  \mathrm{u}\left(
\epsilon\right)  \left(  p_{1}+\frac{p_{1}^{2}-p_{T}^{2}/2}{\epsilon
+m}\right)  +\mathrm{v}\left(  \epsilon\right)  \left(  p_{1}+\frac{p_{1}%
^{2}-p_{T}^{2}/2}{\epsilon-m}\right)  \right)  \delta\left(  \frac
{\epsilon+p_{1}}{M}-x\right)  \frac{d^{3}\mathbf{p}}{\epsilon}. \label{cr35}%
\end{align}
Further, one can easily check (see Appendix \ref{app4}) that%
\begin{equation}
\Gamma_{1}=\int_{0}^{1}g_{1}\left(  x\right)  dx=\frac{1}{6}\left(
\mathrm{U}+\mathrm{V}\right)  +\frac{1}{3}\left(  \mathrm{U}-\mathrm{V}%
\right)  \tilde{\mu}=\left\langle \mathbb{S}_{z}\right\rangle , \label{rs49}%
\end{equation}
where $\mathrm{U,V}$ are constants (\ref{rs46b}). One can verify the term with
the $\mathrm{U,V}$, which are taken from Tab. \ref{tb3} is equal to the
corresponding term in Tab. \ref{tb2}. In fact this comparison represents a
cross-check that our procedure leading to the SFs is correct. Of course, exact
equality $\Gamma_{1}=\left\langle \mathbb{S}_{z}\right\rangle $\ is valid only
in a simplified notation, where quark charges are replaced by $1$. But in the
analysis which aims to extraction of \ $\left\langle \mathbb{S}_{z}%
\right\rangle $ from the experimentally measured $\Gamma_{1}$ one has to take
into account the corresponding charges.

\section{Proton spin structure}

\label{pss}In this section the obtained results are applied to the description
of proton, assuming its spin $J=1/2$ is generated by the spins and OAMs of the
partons, which the proton consists of. The proton state can be formally
represented by a superposition of the Fock states
\begin{equation}
\Psi=\sum_{q,g}a_{qg}\left\vert \varphi_{1},...\varphi_{n_{q}}\right\rangle
\left\vert \psi_{1},...\psi_{n_{g}}\right\rangle , \label{rs20}%
\end{equation}
where the symbols $q,g$\ \ represent the quark and gluon degrees of freedom.
In a first approximation we ignore possible contribution of the gluons and we
study the states%
\begin{equation}
\Psi=\sum_{q}a_{q}\left\vert \varphi_{1},...\varphi_{n_{q}}\right\rangle ,
\label{rs22}%
\end{equation}
where the many-quark states $\left\vert \varphi_{1},...\varphi_{n_{q}%
}\right\rangle $ are represented by the eigenstates $J,J_{z}$ (\ref{rs10c}):%
\begin{equation}
J=J_{z}=\left\langle \mathbb{L}_{z}\right\rangle +\left\langle \mathbb{S}%
_{z}\right\rangle =\frac{1}{2}. \label{rs20a}%
\end{equation}
These states are understood in the context of Sec. \ref{model}, which means
the quarks are considered effectively free only during a short time interval
necessary for the photon absorption. The spin contribution $\left\langle
\mathbb{S}\right\rangle $\ of each many-quark state to the proton spin is
defined by the corresponding matrix element (\ref{rs16}) or equivalently by
the spin vector (\ref{rs47c}), where the scalar functions $\mathrm{u}%
,\mathrm{v}$ depend on the quark energy $\left(  \epsilon=p\cdot P/M\right)  $
and on the pattern of the AM composition. An example of the latter dependence
for $n_{q}=3$ is given in Tabs. \ref{tb2}, \ref{tb3} and the similar tables
could be presented also for the higher $n_{q}=5,7,9,..$. The tables $n_{q}=3$
correspond to the scenario when the spin contribution of the sea quarks is
neglected so the proton spin is generated by the three valence quarks only.
However regardless of $n_{q}$ the corresponding spin SFs are for $J=1/2$
represented by the relations (\ref{cr34}) and (\ref{cr35}).\ \ 

These SFs can be compared with our previous results
\cite{Zavada:2007ww,Zavada:2002uz,Zavada:2001bq}. First, one can observe the
new SFs are identical to the old ones for $\mathrm{v}\left(  \epsilon\right)
=0$. Apparently in this case the new function $\mathrm{u}\left(
\epsilon\right)  $ can be identified with the former phenomenological
distributions $H$ (or $\Delta G$). As before, one can also easily prove (see
Appendix \ref{app4}) Burkhardt-Cottingham sum rule:%
\begin{equation}
\Gamma_{2}=\int_{0}^{1}g_{2}\left(  x\right)  dx=0, \label{rs50}%
\end{equation}
which holds for any $\mathrm{u},\mathrm{v}$. Next, if one assumes massless
quarks, $m\rightarrow0$, then%
\begin{align}
g_{1}\left(  x\right)   &  =\frac{1}{2}\int\left(  \mathrm{u}\left(
\epsilon\right)  +\mathrm{v}\left(  \epsilon\right)  \right)  \left(
p_{1}+\frac{p_{1}^{2}}{\epsilon}\right)  \delta\left(  \frac{\epsilon+p_{1}%
}{M}-x\right)  \frac{d^{3}\mathbf{p}}{\epsilon},\label{rs51}\\
g_{2}\left(  x\right)   &  =-\frac{1}{2}\int\left(  \mathrm{u}\left(
\epsilon\right)  +\mathrm{v}\left(  \epsilon\right)  \right)  \left(
p_{1}+\frac{p_{1}^{2}-p_{T}^{2}/2}{\epsilon}\right)  \delta\left(
\frac{\epsilon+p_{1}}{M}-x\right)  \frac{d^{3}\mathbf{p}}{\epsilon}.
\label{rs52}%
\end{align}
and the sum $\mathrm{u}\left(  \epsilon\right)  +\mathrm{v}\left(
\epsilon\right)  $ can be identified with the former distribution $H\left(
\epsilon\right)  $. It follows that the functions (\ref{rs51}) and
(\ref{rs52}) satisfy the Wanzura-Wilczek (WW), Efremov-Leader-Teryaev (ELT)
and other rules\ that we proved \cite{Zavada:2002uz} for massless quarks. Also
the transversity \cite{Efremov:2004tz} and TMDs
\cite{Efremov:2010mt,Efremov:2009ze} relations keep to be valid. The following
rules are known to be well compatible with the data:

\textit{i)} The Burkhardt-Cottingham integral (\ref{rs50}) has been evaluated
by the experiments \cite{Airapetian:2011wu, Anthony:2002hy,Abe:1998wq}.

\textit{ii) }The ELT sum rule was confirmed in the experiment
\cite{Anthony:2002hy}.

\textit{iii)} The WW relation for the $g_{2}$ SF is compatible with available
data from the experiments \cite{Airapetian:2011wu, Anthony:2002hy,Abe:1998wq}.
Apart from the CQM with massless quarks its validity follows also from the
further approaches \cite{D'Alesio:2009kv,Jackson:1989ph} that are based on the
Lorentz invariance. \ The possible breaking of the WW and other so-called
Lorentz invariance relations were discussed in
\cite{Accardi:2009au,Metz:2008ib}. In our approach this relation is violated
by the mass term, which can be extracted from Eqs. (\ref{rs34}) and
(\ref{rs35}).

However, the most important result of the present paper is related to the
problem of proton spin content $\left\langle \mathbb{S}_{z}\right\rangle =$
$\Delta\Sigma/2.$ Our present calculation again strongly suggest the important
role of the quark OAM in the proton spin. The spin contribution $\left\langle
\mathbb{S}_{z}\right\rangle $ depends on the parameter $\tilde{\mu
}=\left\langle m/\epsilon\right\rangle $ and for a "ground state"
configuration
\begin{equation}
\jmath_{1}=\jmath_{2}=\jmath_{3}=...=\jmath_{n_{q}}=\frac{1}{2}\label{rs20b}%
\end{equation}
we have according to (\ref{RS23})%
\begin{equation}
\left\langle \mathbb{S}_{z}\right\rangle =\frac{1+2\tilde{\mu}}{6}%
,\label{rs26}%
\end{equation}
which for massless quarks, $\tilde{\mu}\rightarrow0$, gives%
\begin{equation}
\left\langle \mathbb{S}_{z}\right\rangle =\frac{1}{6}.\label{rs26a}%
\end{equation}
If there is an admixture of states $j_{k}>1/2$, then one can expect the
condition (\ref{rs19}) is satisfied (provided there is no a~priori preference
among the composition patterns, see paragraph \ref{3fermions}). It means that
\begin{equation}
\Delta\Sigma\lesssim1/3\label{rs27}%
\end{equation}
and the "missing" part of the proton spin is compensated by the quark OAM. The
equivalent result follows from the first moment $\Gamma_{1}$ of the
corresponding SF (\ref{rs49}), from which the $\Delta\Sigma$\ is extracted.
Recent analysis of \ the results from the experiment COMPASS
\cite{Alekseev:2010ub} gives%
\[
\Delta\Sigma=0.32\pm0.03(stat.)
\]
at $Q^{2}=3GeV^{2}/c^{2}$. This result is fully compatible with the former
precision data from the experiments COMPASS and HERMES
\cite{Alexakhin:2006vx,Airapetian:2007mh}. It is obvious this experimental
result agrees very well with the relations (\ref{rs26a}) or (\ref{rs27}),
which have been based on the assumption that the gluon contribution to the
proton spin can be neglected. Such assumption is compatible with the present
experimental estimates \cite{Adolph:2012vj,Airapetian:2010ac}.

The discussion up to now has been devoted to the proton spin SFs. However the
form of the functions (\ref{cr34}),(\ref{cr35}) or (\ref{rs51}),(\ref{rs52})
can be applied to any subset of quarks, for which we can specify corresponding
spin vector. For instance, from the SU(6) approach mentioned on the end of
paragraph \ref{3fermions} one could obtain the spin vectors corresponding to
the $u$ and $d$ quarks and calculate the related spin SFs $g_{1}^{a}$ and
$g_{2}^{a}$, $a=u,d$ and then get the results%
\begin{equation}
\left\langle \mathbb{S}_{z}^{u}\right\rangle =\Gamma_{1}^{u}=\frac{4}%
{3}\left\langle \mathbb{S}_{z}\right\rangle ,\qquad\left\langle \mathbb{S}%
_{z}^{d}\right\rangle =\Gamma_{1}^{d}=-\frac{1}{3}\left\langle \mathbb{S}%
_{z}\right\rangle , \label{rs30}%
\end{equation}
where $\left\langle \mathbb{S}_{z}\right\rangle $ is the full spin
(\ref{rs26}) or (\ref{rs26a}). However, the SU(6) is only an example and rough
approximation. The invariant functions $g_{1}^{u,d}$ have in the IMF, where
the Bjorken $x$\ can be replaced by the light cone ratio, a standard
interpretation of distribution functions.

The basis for obtaining the above predictions related to $g_{1}$ and $g_{2}$
is the covariant description of DIS in which the 3D kinematics is essential.
This is the basic difference from the conventional collinear approach, where
consequently the similar predictions cannot be obtained. Actually the
collinear approach nor allow us to consistently express the function $g_{2}$
\cite{ael}.

\section{ Summary and conclusion}

\label{summary}We have studied the interplay between the spins and OAMs of the
quarks, which are in conditions of DIS effectively free and collectively
generate the proton spin. The basis of this study is the CQM approach
suggested in Sec.\ref{model}. The covariant kinematics is an important
condition for a consistent handling of the OAM. At the same time it is obvious
that the proton rest frame is the proper starting frame for the study of this
interplay. The composition of the contributions from single quarks is defined
by the general rules of AM composition. We have shown the ratio of the quark
effective mass and its energy in the proton rest frame $\tilde{\mu
}=\left\langle m/\epsilon\right\rangle $ plays a crucial role, since it
controls a "contraction" of the spin component which is compensated by the
OAM. Let us point out this effect is a pure consequence of relativistic
kinematics, which does not contradict the fact that the effective quantities
$m$ and $\epsilon$ or their distributions follow from the QCD. In fact the
proton studied at polarized DIS is an ideal instrument for the study of this
relativistic effect. We have shown that the resulting quark spin vector
obtained from composition of the spins of all contributing quarks is a
quantity of key importance. The general form of this vector is given by Eq.
(\ref{rs47c}) and its manifestly covariant representation by Eqs.
(\ref{cr15})$-$(\ref{rs63}). This vector is a basic input for calculation of
the proton spin content and the related SFs. The obtained form of the spin
vector is related to a particle with spin $J=1/2.$ For example the spin vector
corresponding to some baryons with $J=3/2$ would in Eq. (\ref{rs48})\ involve
an additional term proportional to $\cos4\theta,$ cf. related terms $j=3/2$ in
Tab. \ref{tb1}. A very good agreement with the data particularly as for the
$\Delta\Sigma$ is a strong argument in favour of the CQM.

The open question is how the functions $\mathrm{u}\left(  \epsilon
,Q^{2}\right)  ,\mathrm{v}\left(  \epsilon,Q^{2}\right)  $ defining the spin
vector $w$ depend on the scale $Q^{2}$? Is this task calculable in terms of
the perturbative QCD? Another open problem could be related to the method of
experimental measuring of the integral $\mathrm{V}\left(  Q^{2}\right)  $
defined in (\ref{rs46b}). Its nonzero value is related to the possible
admixture of the quark states with $j>1/2$ or $l_{p}\geq1$ in the many-quark
state $J=1/2$.

\begin{acknowledgments}
This work was supported by the project LG130131 of Ministry of Education,
Youth and Sports of the Czech Republic. I am grateful to Anatoli Efremov, Oleg
Teryaev and Peter Schweitzer for many useful discussions and valuable comments.
\end{acknowledgments}

\appendix

\section{Comments on relations ({\ref{RS11G}}), ({\ref{RS23}}) and
({\ref{RS39}})}

\label{app1} Let the state $\Phi_{c,1/2,1/2}(\mathbf{\omega}_{1}%
,\mathbf{\omega}_{2},..\mathbf{\omega}_{n})$ is composed of the $\left(
n-1\right)  $-fermion state and the one-fermion state with angular moments $J$
and $\ j$ respectively:%

\begin{equation}
\Phi_{c,1/2,1/2}(\mathbf{\omega}_{1},\mathbf{\omega}_{2},..\mathbf{\omega
}_{n-1},\mathbf{\omega})=\sum_{j_{zk}=-j}^{j}\left\langle j,j_{z}%
,J,1/2-j_{z}\left\vert 1/2,1/2\right.  \right\rangle \Phi_{j,j_{z}}\left(
\mathbf{\omega}\right)  \Phi_{J,1/2-j_{z}}\left(  \mathbf{\Omega}\right)  ,
\label{rs11b}%
\end{equation}
where $\mathbf{\Omega=\omega}_{1},\mathbf{\omega}_{2},..\mathbf{\omega}_{n-1}%
$. This state generates the distribution
\begin{align}
P_{c}(\mathbf{\omega}_{1},\mathbf{\omega}_{2},..\mathbf{\omega}_{n-1}%
,\mathbf{\omega})  &  =\sum_{j_{z},j_{z}^{\prime}}\left\langle j,j_{z}%
^{\prime},J,1/2-j_{z}^{\prime}\left\vert 1/2,1/2\right.  \right\rangle
\left\langle j,j_{z},J,1/2-j_{z}\left\vert 1/2,1/2\right.  \right\rangle
\label{rs11c}\\
&  \times\Phi_{j,j_{z}^{\prime}}^{+}\left(  \mathbf{\omega}\right)
\Phi_{j,j_{z}}\left(  \mathbf{\omega}\right)  \Phi_{J,1/2-j_{z}^{\prime}}%
^{+}\left(  \mathbf{\Omega}\right)  \Phi_{J,1/2-j_{z}}\left(  \mathbf{\Omega
}\right)  ,\nonumber
\end{align}
where only $J=j\pm1/2$ is allowed due to the triangle condition (\ref{rs9a}).
One can check (\textit{obtained with WM}) the relation%
\begin{equation}
\left\langle j,j_{z},J,1/2-j_{z}\left\vert 1/2,1/2\right.  \right\rangle
^{2}=\frac{1}{2j+1}(1+aj_{z}), \label{rs11d}%
\end{equation}
where $a=1/j$ for $J=j-1/2$ and $a=-1/\left(  j+1\right)  $ for $J=j+1/2$.
Then integration over degrees of freedom $d\mathbf{\Omega=}%
{\displaystyle\prod\limits_{i=1}^{n-1}}
d\mathbf{\omega}_{i}$\ gives a one-fermion distribution%
\begin{equation}
p_{c}(\mathbf{\omega})=\frac{1}{2j+1}\sum_{j_{z}=-j}^{j}(1+aj_{z}%
)\Phi_{j,j_{z}}^{+}\left(  \mathbf{\omega}\right)  \Phi_{j,j_{z}}\left(
\mathbf{\omega}\right)  . \label{rs11e}%
\end{equation}
The terms proportional to $\pm j_{z}$ cancel out and then due to the general
rule%
\begin{equation}
\sum_{j_{z}=-j}^{j}\Phi_{j,j_{z}}^{+}\left(  \mathbf{\omega}\right)
\Phi_{j,j_{z}}\left(  \mathbf{\omega}\right)  =\frac{2j+1}{4\pi} \label{rs11f}%
\end{equation}
the Eq. ({\ref{RS11G}}) follows immediately.

In a similar way one can treat with the distribution ({\ref{RS39}}) and obtain
the form similar to Eq. (\ref{rs11e}):
\begin{equation}
h_{cjl_{p}}(\mathbf{\omega})=\frac{1}{2j+1}\sum_{j_{z}=-j}^{j}(1+aj_{z}%
)\Phi_{jl_{p}j_{z}}^{+}\left(  \mathbf{\omega}\right)  \Delta\mathcal{P}%
\Phi_{jl_{p}j_{z}}\left(  \mathbf{\omega}\right)  . \label{rs11g}%
\end{equation}
This distribution, which is generated by the state ({\ref{rs11b}}) can be
simplified to the form ({\ref{rs41}}), ({\ref{rs40}}) with the factors
$\alpha,\beta$ listed in Tab. \ref{ta1} (\textit{obtained with WM}).
\begin{table}[ptb]
\begin{center}%
\begin{tabular}
[c]{c|c|c|}
& $l=j-1/2$ & $l=j+1/2$\\\hline
$J=j-1/2$ & $\frac{j+3/2}{4j},-\frac{j-1/2}{4j}$ & $-\frac{j-1/2}{4j}%
,\frac{j+3/2}{4j}$\\
$J=j+1/2$ & $-\frac{j+3/2}{4\left(  j+1\right)  },\frac{j-1/2}{4\left(
j+1\right)  }$ & $\frac{j-1/2}{4\left(  j+1\right)  },-\frac{j+3/2}{4\left(
j+1\right)  }$%
\end{tabular}
\end{center}
\caption{The factors $\alpha,\beta$ in distribution (\ref{rs40}) generated by
the composion ({\ref{rs11b}}). }%
\label{ta1}%
\end{table}

The spin contribution ({\ref{rs16}}) can be for $j_{i}=j$ expanded%
\begin{equation}
\left\langle \mathbb{S}_{z}\right\rangle =%
{\displaystyle\sum\limits_{j_{z1}+j_{z2}+...j_{zn}=1/2}}
c_{j}^{2}\left(  \left\langle \left\langle s_{z}\right\rangle \right\rangle
_{j,j_{z1}}+\left\langle \left\langle s_{z}\right\rangle \right\rangle
_{j,j_{z2}}+...+\left\langle \left\langle s_{z}\right\rangle \right\rangle
_{j,j_{zn}}\right)  , \label{rs28}%
\end{equation}
where $c_{j}$ are coefficients ({\ref{rs10b}}). With the use of relation
({\ref{rs8a}}) one gets ({\ref{RS23}}):%
\begin{equation}
\left\langle \mathbb{S}_{z}\right\rangle =%
{\displaystyle\sum\limits_{j_{z1}+j_{z2}+...+j_{zn}=1/2}}
c_{j}^{2}\left(  j_{z1}+j_{z2}+...+j_{zn}\right)  =\frac{1+\left(
2j+1\right)  \tilde{\mu}}{8j\left(  j+1\right)  }. \label{rs29}%
\end{equation}

\section{Spin vector in covariant representation}

\label{app2}The quark spin vector (\ref{cr15}) after contracting with
$P_{\sigma},S_{\sigma},w_{\sigma}$ satisfies the equations%
\begin{align}
AM^{2}+C\left(  p\cdot P\right)   &  =P\cdot w,\label{cr1}\\
-B+C\left(  p\cdot S\right)   &  =S\cdot w,\label{cr2}\\
A\left(  p\cdot P\right)  +B\left(  p\cdot S\right)  +Cm^{2}  &  =0.
\label{cr3}%
\end{align}
At the same time the spin vector $w$ in the quark rest frame reads%
\begin{equation}
w=\left(  0,\mathbf{w}\right)  , \label{cr4}%
\end{equation}
where $\mathbf{w}$ is given by Eq. (\ref{rs47c}). This vector can be
transformed from the quark rest frame to the proton rest frame (where
$P=\left(  M,0,0,0\right)  $ and $S=\left(  0,\mathbf{S}\right)  $). After
decomposition of the vector $\mathbf{w}$ to longitudinal and transversal parts
with respect to the quark momentum $\mathbf{p}$ in the proton rest frame, the
corresponding Lorentz boost gives%
\begin{equation}
(0,\mathbf{w})\rightarrow w=\left(  \frac{\mathbf{p\cdot w}}{m},\,\mathbf{w}%
+\frac{\mathbf{p\cdot w}}{m(\epsilon+m)}\mathbf{p}\right)  . \label{cr12}%
\end{equation}
One can check that%
\begin{equation}
\mathbf{p\cdot w}=-\left(  p\cdot S\right)  \left(  \mathrm{u}+\mathrm{v}%
\right)  , \label{cr11}%
\end{equation}
then substitution to (\ref{cr12}) gives%
\begin{align}
P\cdot w  &  =-\frac{M}{m}\left(  p\cdot S\right)  \left(  \mathrm{u}%
+\mathrm{v}\right)  ,\label{cr10}\\
S\cdot w  &  =-\left(  \mathrm{u}-\mathrm{v}\right)  -\frac{\left(  p\cdot
S\right)  ^{2}}{m}\left(  \frac{\mathrm{u}}{\epsilon+m}+\frac{\mathrm{v}%
}{\epsilon-m}\right)  . \label{cr9}%
\end{align}
One can check the equations (\ref{cr1})$-$(\ref{cr3}) after substitution from
(\ref{cr10}),(\ref{cr9}) give solution (\ref{rs61})$-$(\ref{rs63}).

\section{Spin structure functions and proton rest frame}

\label{app3}The integrals (\ref{cr29}) are calculated similarly as in Appendix
of the paper \cite{Zavada:2001bq}. For integration we use the proton rest
frame in which
\begin{equation}
\mathbf{p}=p_{1}\mathbf{e}_{1}+p_{2}\mathbf{e}_{2}+p_{3}\mathbf{e}_{3}%
,\quad\mathbf{e}_{1}=-\frac{\mathbf{q}}{\left\vert \mathbf{q}\right\vert
},\quad\mathbf{e}_{2}=\frac{\mathbf{S}-(\mathbf{S\cdot e}_{1})\mathbf{e}_{1}%
}{\sqrt{1-(\mathbf{S\cdot e}_{1})^{2}}},\quad\mathbf{e}_{3}=\mathbf{e}%
_{1}\times\mathbf{e}_{2}, \label{asf1}%
\end{equation}
so one gets
\begin{equation}
\mathbf{p\cdot q}=-p_{1}\left\vert \mathbf{q}\right\vert ,\qquad\mathbf{p\cdot
S}=-p_{1}\cos\zeta+p_{2}\sin\zeta,\qquad\cos\zeta\equiv\frac{\mathbf{q\cdot
S}}{\left\vert \mathbf{q}\right\vert }. \label{asf2}%
\end{equation}
In this reference frame, for $Q^{2}\gg4M^{2}x^{2}$ we have $\left\vert
\mathbf{q}\right\vert /\nu\rightarrow1$ (see e.g. \cite{Zavada:2011cv}), which
gives:%
\begin{align}
P\cdot w  &  =\frac{M}{m}\left(  \mathbf{p\cdot S}\right)  \left(
\mathrm{u}+\mathrm{v}\right)  ,\label{asf3}\\
S\cdot w  &  =-\left(  \mathrm{u}-\mathrm{v}\right)  -\frac{\left(
\mathbf{p\cdot S}\right)  ^{2}}{m}\left(  \frac{\mathrm{u}}{\epsilon+m}%
+\frac{\mathrm{v}}{\epsilon-m}\right)  ,\label{asf4}\\
q\cdot w  &  =\nu\left(  \left(  \mathbf{p\cdot S}\right)  \left(
\frac{\mathrm{u}}{\epsilon+m}-\frac{\mathrm{v}}{\epsilon-m}\right)  -\left(
\mathrm{u}-\mathrm{v}\right)  \cos\zeta+\frac{\mathbf{p\cdot S}}{m}\left(
\frac{\mathrm{u}}{\epsilon+m}+\frac{\mathrm{v}}{\epsilon-m}\right)  \left(
\epsilon+p_{1}\right)  \right)  ,\label{asf5}\\
D  &  =-\frac{\nu\left\{  P\cdot w\right\}  /M+\nu\left\{  S\cdot w\right\}
\cos\zeta-\left\{  q\cdot w\right\}  }{\nu^{2}\sin^{2}\zeta},\label{asf6}\\
\delta\left(  \frac{p\cdot q}{P\cdot q}-x\right)   &  =\delta\left(
\frac{\epsilon+p_{1}}{M}-x\right)  \label{asf7}%
\end{align}
These terms, after substitution to Eqs. (\ref{cr26}),(\ref{cr27}) allow us to
calculate the integrals $G_{P},G_{S}$ and their combinations (\ref{cra31})
giving $g_{1}$ and $g_{2}$. After substitution
\[
p_{2}=p_{T}\cos\varphi,\quad p_{3}=p_{T}\sin\varphi,\quad d^{3}\mathbf{p}%
=p_{T}dp_{T}dp_{1}d\varphi
\]
and integration over $\varphi$ we get:%
\begin{align}
g_{1}\left(  x\right)   &  =\pi\int\left(  \mathrm{u}\left(  \epsilon\right)
\left(  p_{1}+m+\frac{p_{1}^{2}}{\epsilon+m}\right)  +\mathrm{v}\left(
\epsilon\right)  \left(  p_{1}-m+\frac{p_{1}^{2}}{\epsilon-m}\right)  \right)
\delta\left(  \frac{\epsilon+p_{1}}{M}-x\right)  \frac{p_{T}dp_{T}dp_{1}%
}{\epsilon},\label{asf9}\\
g_{2}\left(  x\right)   &  =-\pi\int\left(  \mathrm{u}\left(  \epsilon\right)
\left(  p_{1}+\frac{p_{1}^{2}-p_{T}^{2}/2}{\epsilon+m}\right)  +\mathrm{v}%
\left(  \epsilon\right)  \left(  p_{1}+\frac{p_{1}^{2}-p_{T}^{2}/2}%
{\epsilon-m}\right)  \right)  \delta\left(  \frac{\epsilon+p_{1}}{M}-x\right)
\frac{p_{T}dp_{T}dp_{1}}{\epsilon}. \label{asf10}%
\end{align}
Then the substitution $\ 2\pi p_{T}dp_{T}dp_{1}=d^{3}\mathbf{p}$ gives the
relations (\ref{cr26}) and (\ref{cr27}).

\section{First moments}

\label{app4}Eq. (\ref{rs51}) implies%
\[
\Gamma_{1}=\int_{0}^{1}g_{1}\left(  x\right)  dx=\frac{1}{2}\int\left(
\mathrm{u}\left(  \epsilon\right)  \left(  p_{1}+m+\frac{p_{1}^{2}}%
{\epsilon+m}\right)  +\mathrm{v}\left(  \epsilon\right)  \left(  p_{1}%
-m+\frac{p_{1}^{2}}{\epsilon-m}\right)  \right)  \frac{d^{3}\mathbf{p}%
}{\epsilon}.
\]
Due to rotational symmetry the integral simplifies%
\[
\Gamma_{1}=\frac{1}{2}\int\left(  \mathrm{u}\left(  \epsilon\right)  \left(
m+\frac{\mathbf{p}^{2}/3}{\epsilon+m}\right)  +\mathrm{v}\left(
\epsilon\right)  \left(  -m+\frac{\mathbf{p}^{2}/3}{\epsilon-m}\right)
\right)  \frac{d^{3}\mathbf{p}}{\epsilon}%
\]
and taking into account that $\mathbf{p}^{2}=\left(  \epsilon+m\right)
\left(  \epsilon-m\right)  $ $\ $we obtain%
\begin{align*}
\Gamma_{1}  &  =\frac{1}{6}\int\left(  \mathrm{u}\left(  \epsilon\right)
+\mathrm{v}\left(  \epsilon\right)  +\frac{2m}{\epsilon}\left(  \mathrm{u}%
\left(  \epsilon\right)  -\mathrm{v}\left(  \epsilon\right)  \right)  \right)
d^{3}\mathbf{p}\\
&  =\frac{1}{6}\left(  \mathrm{U}+\mathrm{V}\right)  +\frac{1}{3}\left(
\mathrm{U}-\mathrm{V}\right)  \tilde{\mu}.
\end{align*}
In a similar way with the use of rotational symmetry, one can prove also Eq.
(\ref{rs50}).


\begin{thebibliography}{99}                                                                                               %




\bibitem {Aidala:2012mv}C.~A.~Aidala, S.~D.~Bass, D.~Hasch and G.~K.~Mallot,
Rev.\ Mod.\ Phys.\ \textbf{85}, 655 (2013) [arXiv:1209.2803 [hep-ph]].



\bibitem {Myhrer:2009uq}F.~Myhrer and A.~W.~Thomas,
J.\ Phys.\ G \textbf{37}, 023101 (2010) [arXiv:0911.1974 [hep-ph]].



\bibitem {Burkardt:2008jw}M.~Burkardt, C.~A.~Miller and W.~D.~Nowak,
Rept.\ Prog.\ Phys.\ \textbf{73}, 016201 (2010) [arXiv:0812.2208 [hep-ph]].



\bibitem {Barone:2010zz}V.~Barone, F.~Bradamante and A.~Martin,
Prog.\ Part.\ Nucl.\ Phys.\ \textbf{65}, 267 (2010) [arXiv:1011.0909 [hep-ph]].



\bibitem {Kuhn:2008sy}S.~E.~Kuhn, J.~-P.~Chen and E.~Leader,
Prog.\ Part.\ Nucl.\ Phys.\ \textbf{63}, 1 (2009) [arXiv:0812.3535 [hep-ph]].



\bibitem {Zavada:2011cv}P.~Zavada,
Phys.\ Rev.\ D \textbf{85}, 037501 (2012) [arXiv:1106.5607 [hep-ph]].

\bibitem {Zavada:2009sk}P.~Zavada, Phys.\ Rev.\ D \textbf{83}, 014022 (2011)
[arXiv:0908.2316 [hep-ph]].


\bibitem {Zavada:2007ww}P.~Zavada,
Eur.\ Phys.\ J.\ C \textbf{52}, 121 (2007).


\bibitem {Zavada:2002uz}P.~Zavada,
Phys.\ Rev.\ D \textbf{67}, 014019 (2003).


\bibitem {Zavada:2001bq}P.~Zavada,
Phys.\ Rev.\ D \textbf{65}, 054040 (2002).


\bibitem {Zavada:1996kp}P.~Zavada,
Phys.\ Rev.\ D \textbf{55}, 4290 (1997).


\bibitem {Efremov:2010cy}A.~V.~Efremov, P.~Schweitzer, O.~V.~Teryaev and
P.~Zavada,
PoS \textbf{DIS2010}, 253 (2010) [arXiv:1008.3827 [hep-ph]].


\bibitem {Efremov:2010mt}A.~V.~Efremov, P.~Schweitzer, O.~V.~Teryaev and
P.~Zavada,
Phys.\ Rev.\ D \textbf{83}, 054025 (2011).


\bibitem {Efremov:2009ze}A.~V.~Efremov, P.~Schweitzer, O.~V.~Teryaev and
P.~Zavada,
Phys.\ Rev.\ D \textbf{80}, 014021 (2009) [arXiv:0903.3490 [hep-ph]].


\bibitem {Efremov:2004tz}A.~V.~Efremov, O.~V.~Teryaev and P.~Zavada,
Phys.\ Rev.\ D \textbf{70}, 054018 (2004).


\bibitem {wolfram}Wolfram Research, Inc., \textit{Mathematica,} Version 9.0,
Champaign, IL (2011).

\bibitem {fey}R. P. Feynman, \textit{Photon-Hadron Interactions, }Benjamin,
New York, 1972.

\bibitem {ioffe}B.L. Ioffe, Phys. Lett. B \textbf{30,} 129 (1969).

\bibitem {imf}%
\texttt{http://en.wikipedia.org/wiki/Parton\_(particle\_physics)}

\bibitem {bdtm}V.B. Berestetsky, A.Z. Dolginov and K.A. Ter-Martirosyan,
\textit{Angular Functions of Particles with Spins }(in Russian), JETP
\textbf{20,} 527-535 (1950).

\bibitem {lali}L.D. Landau, E.M. Lifshitz \textit{et al.}, \textit{Quantum
Electrodynamics} (Course of Theoretical Physics, vol. 4), Elsevier Science
Ltd., 1982.

\bibitem {bie}L.C. Biedenharn, J.D. Louck, \textit{Angular Momentum in Quantum
Physics: Theory and Application}, Cambridge University Press 1985.

\bibitem {D'Alesio:2009kv}U.~D'Alesio, E.~Leader and F.~Murgia,
Phys.\ Rev.\ D \textbf{81}, 036010 (2010) [arXiv:0909.5650 [hep-ph]].


\bibitem {Jackson:1989ph}J.~D.~Jackson, G.~G.~Ross and R.~G.~Roberts,
Phys.\ Lett.\ B \textbf{226}, 159 (1989).




\bibitem {Abe:1998wq}K.~Abe \textit{et al.} [E143 Collaboration],
Phys.\ Rev.\ D \textbf{58}, 112003 (1998) [hep-ph/9802357].




\bibitem {Anthony:2002hy}P.~L.~Anthony \textit{et al.} [E155 Collaboration],
Phys.\ Lett.\ B \textbf{553}, 18 (2003) [hep-ex/0204028].




\bibitem {Accardi:2009au}A.~Accardi, A.~Bacchetta, W.~Melnitchouk and
M.~Schlegel,
JHEP \textbf{0911}, 093 (2009) [arXiv:0907.2942 [hep-ph]].



\bibitem {Metz:2008ib}A.~Metz, P.~Schweitzer and T.~Teckentrup,
Phys.\ Lett.\ B \textbf{680}, 141 (2009) [arXiv:0810.5212 [hep-ph]].




\bibitem {Airapetian:2011wu}A.~Airapetian, N.~Akopov, Z.~Akopov,
E.~C.~Aschenauer, W.~Augustyniak, R.~Avakian, A.~Avetissian and E.~Avetisyan
\textit{et al.},
Eur.\ Phys.\ J.\ C \textbf{72}, 1921 (2012) [arXiv:1112.5584 [hep-ex]].



\bibitem {Alekseev:2010ub}M.~G.~Alekseev \textit{et al.} [COMPASS
Collaboration],
Phys.\ Lett.\ B \textbf{693}, 227 (2010) [arXiv:1007.4061 [hep-ex]].




\bibitem {Alexakhin:2006vx}V.~Y.~.Alexakhin \textit{et al.} [COMPASS
Collaboration],
Phys.\ Lett.\ B \textbf{647}, 8 (2007) [hep-ex/0609038].




\bibitem {Airapetian:2007mh}A.~Airapetian \textit{et al.} [HERMES
Collaboration],
Phys.\ Rev.\ D \textbf{75}, 012007 (2007) [hep-ex/0609039].




\bibitem {Adolph:2012vj}C.~Adolph \textit{et al.} [COMPASS Collaboration],
Phys.\ Lett.\ B \textbf{718}, 922 (2013) [arXiv:1202.4064 [hep-ex]].



\bibitem {Airapetian:2010ac}A.~Airapetian \textit{et al.} [HERMES
Collaboration],
JHEP \textbf{1008}, 130 (2010) [arXiv:1002.3921 [hep-ex]].

\bibitem {ael}M. Anselmino, A. Efremov, and E. Leader, Phys. Rep.
\textbf{261}, 1 (1995).
\end{thebibliography}
\end{document}